\documentclass[aps,pra,superscriptaddress,floatfix,showpacs,twocolumn]{revtex4-1}

\usepackage{amsmath,amsfonts,amssymb}
\usepackage{graphicx, tikz}
\usepackage{algorithm}
\usepackage{algorithmic}
\usepackage{epstopdf}
\usepackage{mathbbol,bbm}
\usepackage{natbib}

\let\originalleft\left
\let\originalright\right
\renewcommand{\left}{\mathopen{}\mathclose\bgroup\originalleft}
\renewcommand{\right}{\aftergroup\egroup\originalright}
\newcommand{\bra}[1]{\ensuremath{\left\langle #1\right|}}
\newcommand{\ket}[1]{\ensuremath{\left|#1\right\rangle}}

\newcommand{\ie}{\emph{i.e.}}
\newcommand{\eg}{\emph{e.g.}}

\newcommand{\braket}[2]{\left\langle #1,#2\right\rangle}

\renewcommand{\>}{\rangle}
\providecommand{\abs}[1]{\left\lvert#1\right\rvert}
\providecommand{\norm}[1]{\lVert#1\rVert}

\begin{document}
\title{A Statistical Theory of Designed Quantum Transport Across Disordered Networks}
\author{Mattia Walschaers} 
\email{mattia@itf.fys.kuleuven.be}
\affiliation{Physikalisches Institut, Albert-Ludwigs-Universit\"at Freiburg, Hermann-Herder-Str. 3, D-79104 Freiburg, Germany}
\affiliation{Instituut voor Theoretische Fysica, University of Leuven, Celestijnenlaan 200D, B-3001 Heverlee, Belgium}
\author{Roberto Mulet} 
\email{roberto.mulet@gmail.com}
\affiliation{Physikalisches Institut, Albert-Ludwigs-Universit\"at Freiburg, Hermann-Herder-Str. 3, D-79104 Freiburg, Germany}
\affiliation{Complex Systems Group, Department of Theoretical Physics, University of Havana, Cuba}
\author{Thomas Wellens} 
\email{thomas.wellens@physik.uni-freiburg.de}
\affiliation{Physikalisches Institut, Albert-Ludwigs-Universit\"at Freiburg, Hermann-Herder-Str. 3, D-79104 Freiburg, Germany}
\author{Andreas Buchleitner}
\email{a.buchleitner@physik.uni-freiburg.de}
\affiliation{Physikalisches Institut, Albert-Ludwigs-Universit\"at Freiburg, Hermann-Herder-Str. 3, D-79104 Freiburg, Germany}
\affiliation{Freiburg Institute for Advanced Studies, Albert-Ludwigs-Universit\"at Freiburg, Albertstr. 19, D-79104 Freiburg, Germany}

\date{\today}

\begin{abstract}
We explain how centrosymmetry, together with a dominant doublet of energy eigenstates in the local density of states, can guarantee interference-assisted,  strongly enhanced, strictly coherent quantum excitation 
transport between two predefined sites of a random network of two-level systems. Starting from a generalisation of the 
{\em chaos assisted tunnelling} mechanism, we formulate a random matrix theoretical framework for the analytical prediction 
of the transfer time distribution, of lower bounds of the transfer efficiency, and of the scaling behaviour of characteristic statistical properties
with the size of the network. We show that these analytical predictions compare well to numerical simulations, using Hamiltonians sampled from the Gaussian Orthogonal Ensemble (GOE).
\end{abstract}

\maketitle

\section{Introduction}

The impact of quantum interference effects on transport phenomena defines a multi-facetted area of research, with a wide range of 
incarnations in condensed \cite{Ashcroft:1988aa} and soft matter \cite{dubin2005}, mesoscopic physics \cite{ImryBook}, quantum 
chaos \cite{Kottos:1999aa,Madronero:2006aa}, quantum 
computing \cite{hein2009,PhysRevLett.92.187902}, light-matter interaction \cite{wiersma1997,labeyrie1999,wellens2009,hartung2008,joerder2014,modugno2010,wimberger2003}, and, rather recently, 
photobiology \cite{Collini:2009aa,Engel:2007aa,Mancal:2012aa,Walschaers:2013aa,ritz2008,scholes2012}. However, the {\em deterministic 
control} of quantum interference contributions to transport is rightfully considered a subtle problem which turns ever 
more difficult with an increasing density of states, since this implies that more and more relative phases need to be carefully controlled.
Any uncontrolled perturbation of these has then potentially very detrimental effects on the control target (much as in a misaligned Fabry-P\'erot cavity \cite{Hercher:1968aa}). This is why quantum 
engineers traditionally {\em dislike} noise and disorder, generally invoking strong symmetry properties (such as the translational 
invariance of a lattice) to guarantee that the desired quantum effects prevail. Of course, as the system size is scaled up, and almost unavoidably so its 
complexity, perturbations of such symmetries get ever more likely.

On the other hand, it has long been known in solid state and statistical physics that quantum interference 
effects can actually induce very strong signatures on the {\em statistics} of characteristic transport coefficients, even in the presence of 
strong disorder -- Anderson localisation arguably being the most prominent example 
\cite{PhysRevLett.42.673,PhysRev.109.1492,modugno2010}. More recently, it therefore emerges in diverse areas that disorder may actually be 
conceived as a robust handle of ({\em statistical} rather than deterministic) 
quantum control \cite{Madronero:2006aa,Scholak2010,Walschaers:2013aa,krueger2013,Hildner:2013aa,Mostarda:2013aa}, in particular on scales which 
preclude deterministic control on a microscopic level.

One possible, specific scenario for such statistical quantum control is motivated by the ever more consolidating experimental evidence for 
non-trivial, long-lasting quantum coherence in the strongly optimised excitation transport in photosynthetic light harvesting complexes of 
plants and bacteria \cite{Collini:2009aa,Engel:2007aa,Mancal:2012aa}. These supra-molecular and hierarchically structured objects come in rather variable architectures for 
different biological species, but all share the functional purpose of transporting 
energy
to some reaction centre where the plant chemistry is initiated. Ideally, this energy transport should occur with minimal
loss, and that might be an evolutionary incentive for also {\em rapid} transport. Yet, irrespective of their specific, coarse grained 
architectures, all these complexes are garnished by some level of disorder, i.e. their different realisations in the same biological organism 
exhibit modifications on the microscopic level, simply as a consequence of the enormous complexity of the larger biological structure
they are part of. Therefore, the experimentally documented efficiency (close to $100\%$) of the excitation transport {\em unavoidably} 
implies a disorder average, $\langle e^{-it H} \rangle_{\rm disorder}$ (where $H$ is the Hamiltonian), and tells us that nature found a way to guarantee near-to-deterministic delivery {\em despite} the presence of 
uncontrolled structural variations on a microscopic level. 
This stands against a common practice in the literature \cite{Plenio:2008aa,Mohseni:2008aa}, where one uses published Hamiltonian data, \eg~ \cite{Moix:2011aa}, to describe the coherent backbone dynamics in these molecular complexes: Since these data in general result from (typically spectroscopic) experiments on solutions of such complexes, fluctuations cannot be resolved and an implicit disorder average in the reconstructed Hamiltonian, $\langle H\rangle_{\rm disorder}$, is always present. The dynamics, however, is {\em not} self-averaging, $\langle e^{-it H} \rangle_{\rm disorder} \neq e^{-it\langle H\rangle_{\rm disorder}}$, and therefore using such average Hamiltonians will typically fail to capture all the relevant physics. The philosophy of our present contribution is exactly to emphasise the potential of disorder-induced statistical effects to optimise relevant transport observables, such as the transfer efficiency, in the presence of quantum interference. Ultimately, such approach may help to identify experimentally implementable methods to certify the quantum or rather classical origin of the observed transfer efficiencies.

We did argue earlier \cite{Scholak2010,PhysRevE.83.021912,Walschaers:2013aa,Zech:2014aa} that one possible, and strictly quantum, candidate 
mechanism leading to large and exceptionally
rapid excitation transfer in photosynthetic light harvesting units is constructive multi-path quantum interference of the many 
transmission amplitudes from input to output: Reducing the macromolecular complex to a random network, the 
molecular sub-units which constitute the complex are localised at the network's nodes and considered as identical two-level systems with  
two distinct electronic states, coupled by dipole-dipole interactions. In such strongly simplifying model, the randomness 
of the  network's
sites' positions substitutes for the realisation-dependent changes of the 
local environment of the molecular network's constituents, and accounts for the uncertainties in the matrix representations of the effective 
Hamiltonians which can be found in the literature \cite{Moix:2011aa}. Even though minimalistic, we argue that this description proves to be qualitatively sufficient in capturing the essential physics which arises due to disorder. Clearly, this approach is inspired by the fundamental idea of random 
matrix theory (RMT) \cite{Mehta:2004aa}, and strong, quantum interference-induced 
fluctuations of characteristic transport coefficients are to be expected when sampling over 
different network realisations. We could show \cite{Walschaers:2013aa} that the statistics of these fluctuations can be efficiently controlled by 
imposing just two constraints on the otherwise random structure of the network -- centrosymmetry and the presence of a dominant doublet
in the network's spectrum. With these ingredients, it is indeed possible to make the distribution of transfer efficiencies collapse on a 
narrow interval very close to unity, and to guarantee rather rapid transfer times, {\em without} the need to control the microscopic 
hardwiring of the network -- a clear incident of the above {\em statistical quantum control}.

It is the purpose of the present article to spell out the details of the underlying theory, and to scrutinise the scaling properties of the 
thus ``engineered'' statistical distributions with the network size -- \ie~the number of its elementary molecular sites. Given the 
generality of the random graph model which we are building on, as well as the ubiquity of disorder or structural perturbations in large 
networks, we trust that the results here presented do not only provide a fresh  perspective for the discussion of quantum effects in photosynthetic 
light harvesting, but equally much on excitation transport in cold Rydberg gases \cite{PhysRevA.90.063415}, as well as on quantum walks on random graphs or on robust, quantum walk-based quantum 
computing design \cite{Broadbent:2009aa,PhysRevLett.102.180501,PhysRevLett.113.083602}.

\section{The model}
\label{sec:model}

Consider a single excitation propagating on a disordered network of $N$ sites. 
To each site ``$i$'' we associate a quantum state $\ket{i}$ which represents the state where the excitation is fully localised at this very site. These states 
span the single-excitation Hilbert space of our model. The goal is to transport the excitation from an input site $\ket{{\rm in}}$ to an output site $\ket{{\rm out}}$ \footnote{The requirement that the initial and final states, $\ket{{\rm in}}$ and $\ket{{\rm out}}$, be localised on individual sites, is, however, not a strictly necessary ingredient for our subsequent conclusions.}. To mimic disorder, we describe the interaction among the sites by a $N \times N$  Hamiltonian $H$  chosen from the Gaussian Orthogonal Ensemble (GOE)\cite{Mehta:2004aa}, with the additional constraint that the Hamiltonian be centrosymmetric with respect to $\ket{\rm in}$ and $\ket{\rm out}$.  This symmetry is defined by $JH=HJ$ and $\ket{{\rm in}}=J\ket{{\rm out}}$, where $J$ is the exchange matrix, $J_{i,j}=\delta_{i,N-j+1}$\cite{Cantoni:1976aa}. This {\em design principle} is motivated by previous results \cite{ZECH:2013aa,Zech:2014aa} suggesting that centrosymmetric Hamiltonians deduced from dipoles randomly distributed within a sphere are statistically more likely to mediate efficient transport than unconstrained random Hamiltonians.\\

In technical terms, the GOE is characterised by the parameter $\xi$, which describes the density of states as half the radius of Wigner's semicircle \cite{BohigasLesHouches1989}. More explicitly, we define our  ensemble of interest in terms of a probability distribution on matrix elements given by

 \begin{equation}H_{ij} \sim\begin{cases} \mathcal{N}\left(0, \frac{2 \xi^2}{N}\right)\quad\quad &\text{if } i=j \text{ or } i=N-j+1 \\
\mathcal{N}\left(0, \frac{\xi^2}{N}\right)& \text{else}\end{cases},\label{eq:ProbCGOEEven}
\end{equation} where $\mathcal{N}$ denotes the normal distribution with its mean and variance as first and second argument, respectively. The centrosymmetry constraint practically implies that $H_{i,j}=H_{i,N-j+1}=H_{N-i+1,j}=H_{N-i+1,N-j+1}$ (\ie, the matrix representation of $H$ is invariant under mirroring with respect to the matrix' centre), which also guarantees that $E=H_{\text{in},\text{in}}=H_{\text{out},\text{out}}$.
The choice of a variance $\xi^2/N$ is closely related to the behaviour of the spectral density. The specific scaling with $N$ guarantees that the ensemble averaged density of states is independent of $N$, and is always given by a semicircular distribution of radius $2 \xi$  \cite{BohigasLesHouches1989}. 

Within this ensemble, the input and output sites (and therefore also the associated states) are defined as those that couple the weakest, with coupling $V=\min_i \abs{H_{i, N-i+1}}$. This definition originates from the idea that the input and output are ``farthest apart'' (what is a suggestive assumption, \eg~when considering the paradigmatic Fenna Matthews Olson (FMO) light harvesting complex as a macromolecular, 3D ``wire'' which connects the antenna complex to the reaction center \cite{Moix:2011aa}). To avoid the necessity to distinguish between $H_{\text{in},\text{out}}$ and $V$, we will always consider $H_{\text{in},\text{out}}$ to be positive. This boils down to multiplying the full Hamiltonian by $-1$ if $H_{\text{in},\text{out}}$ is negative for some sampled Hamiltonian. It can be easily verified that this will not cause any problems in the following derivations, yet makes the notation somewhat lighter.\\

Each of the thus defined Hamiltonians generates a time evolution $\ket{\phi (t)}=\exp(-it H)\ket{\phi(0)}$ (we set $\hbar\equiv1$) of the initial state $\ket{\phi (0)}= \ket{{\rm in}}$. Focussing on the excitation transfer from $\ket{\rm in}$ to $\ket{\rm out}$, a possible measure of the transfer efficiency is:
 \begin{equation}
\label{eq:eff}\mathcal{P}_H = \max_{t\in \left[0,T_R\right)} \abs{\braket{\text{out}}{\phi(t)}}^2,
\end{equation}
 where $T_R$ is the Rabi time, given by $T_R=\pi/ 2 V$ \cite{Scholak2010,PhysRevE.83.021912} \footnote{Note that we here employ a reference time which is ten times larger than in \cite{Scholak2010,PhysRevE.83.021912}. However, this does not alter the qualitative result, also see \cite{Scholak:2011aa}.}. This is the time needed for an excitation to be fully transferred from input to output when all sites except for $\ket{{\rm in}}$ and $\ket{{\rm out}}$ are discarded. Therefore, transport can be considered ``efficient'' if the intermediate sites of the network accelerate the transfer process as compared to the direct coupling between $\ket{{\rm in}}$ and $\ket{{\rm out}}$. We thus set out to identify necessary and/or sufficient conditions for $H$ to be efficient, and to render the transport as fast as possible.

\subsection{Centrosymmetry}\label{sec:Centro}

We start with a closer scrutiny of the properties of centrosymmetric matrices, and emphasise those aspects which are relevant in the context of quantum transport theory. We will explain why centrosymmetry is an important design principle to enhance the excitation transfer, and also indicate why this symmetry alone is insufficient to guarantee efficiency in the above sense.\\

Given the definition (\ref{eq:eff}) of ${\cal P}_H$, we are interested in the behaviour of  $\abs{\braket{\text{out}}{\phi(t)}}^2$. To relate transport properties to the spectral properties of the underlying Hamiltonian, we use the spectral decomposition
\begin{equation}
\abs{\braket{\text{out}}{\phi(t)}}^2= \abs{\sum^N_{i=1} e^{-i t E_i} \braket{{\rm out}}{\eta_i} \braket{\eta_i}{{\rm in}}}^2,
\end{equation}
where $\eta_i$ and $E_i$ denote the eigenvectors and eigenvalues of the Hamiltonian $H,$ respectively. This expression highlights the eigenvectors' very crucial role for the transport: They determine which sites can be reached from a given input site. If there were no eigenvectors that are significantly localised on both, $\ket{{\rm in}}$ and $\ket{{\rm out}}$, transport would not be possible. The eigenvalues determine the timescale at which transport occurs. 

Centrosymmetry mainly impacts the eigenvectors of the Hamiltonian: It is shown in \cite{Cantoni:1976aa} that a centrosymmetric matrix also has centrosymmetric eigenvectors. This implies that $J \ket{\eta_i}= \pm \ket{\eta_i}$, where $J$ is the symmetry operator as defined at the beginning of Section \ref{sec:model}. Since we define the Hamiltonian to be centrosymmetric with respect to input and output, we know that, by construction, $J\ket{{\rm out}}= \ket{{\rm in}}$. With the centrosymmetry of the eigenvectors, it follows that $\braket{{\rm out}}{\eta_i} \braket{\eta_i}{{\rm in}} = \pm \abs{\braket{{\rm in}}{\eta_i}}^2= \pm \abs{\braket{{\rm out}}{\eta_i}}^2$. Consequently, there is a relation between the probability to have transport from in to out and the return probability. Since we know that, due to weak localisation effects, there is always an enhanced return probability \cite{Scholak2010,Akkermans2007}, we expect to find a corresponding effect for the transfer from in to out.\\

Due to its centrosymmetry, $H$ can be cast, through an orthogonal transformation, into the following block diagonal representation \cite{Cantoni:1976aa} in the eigenbasis of the exchange matrix $J$:  \begin{equation}\label{eq:Hblock}H = \begin{pmatrix} H^+ &0 \\ 0 & H^-\end{pmatrix}.\end{equation} Both, $H^+$ and $H^-$, are $N/2\times N/2$ matrices from the GOE. This is a consequence of the block diagonalisation \cite{Cantoni:1976aa}, combined with the fact that the sum of normally distributed variables is itself a normally distributed variable.

Two eigenvectors of $J$ have the form \begin{equation}\label{eq:pm}\ket{\pm}=\frac{1}{\sqrt{2}}\left(\ket{\text{in}} \pm \ket{\text{out}}\right).\end{equation} Using $\ket{+}$ and $\ket{-}$ to express $\ket{\text{in}}$ and $\ket{\text{out}}$ allows us to rewrite (\ref{eq:eff}) as \begin{equation}\label{eq:effrewrite}\mathcal{P}_H = \max_{t\in \left[0,T_R\right)} \frac{1}{4}\abs{\<e^{-i t H^+ }\>_+ - \<e^{-i t H^- }\>_-}^2.\end{equation} The two terms in this expression are statistically independent. Hence, we need to understand the evolution of $\ket{+}$ and $\ket{-}$ under the unitaries generated by $H^+$ and $H^-$. 
In order to do so, we express $\mathcal{P}_H$ in terms of the eigenvectors $\ket{\eta^{\pm}_i}$ and of the eigenvalues $E_i^{\pm}$ of $H^{\pm}$:
\begin{equation}\begin{split}
 \mathcal{P}_H= \max_{t\in \left[0,T_R\right)}\frac{1}{4}\abs{\sum_i  e^{-i t E^+_i} \abs{\braket{\eta^+_i}{+}}^2- \sum_i  e^{-i t E^-_i} \abs{\braket{\eta^-_i}{-}}^2}^2.
\label{eq:efflong}
\end{split}
\end{equation}
(\ref{eq:efflong}) is our final result for the transfer efficiency when only assuming centrosymmetry. Since eigenvectors $\ket{\eta_i^{\pm}}$ and eigenvalues $E_i^{\pm}$ are stochastic variables described by random matrix statistics, ${\cal P}_H$ will typically exhibit strong interference effects. While centrosymmetry tends to enhance the transfer efficiency via a mechanism related to weak localisation \cite{Akkermans2007}, it still does not prevent the excitation to spread essentially uniformly over the network, as can be seen from the time averaged output site population 
\begin{equation}
p_H = \lim_{T \rightarrow \infty} \frac{1}{T} \int^T_{0} {\rm d}t ~ \abs{\braket{\text{out}}{\phi(t)}}^2 = \sum^N_{i=1}\abs{\braket{{\rm out}}{\eta_i} \braket{\eta_i}{{\rm in}} }^2.
\end{equation}
Due to centrosymmetry, this can be rewritten as $p_H=\sum^N_{i=1}\abs{\braket{{\rm in}}{\eta_i}}^4$, a quantity closely related to the participation ratio \cite{Haake:2010aa}. From \cite{haake_random-matrix_1990},  one obtains for its ensemble average 
\begin{equation}\label{eq:AverPop}
\overline{p_H} =\frac{3}{2+N}.
\end{equation}
This implies that, on average, at least $N/3$ eigenvectors  (with their associated eigenvalues) contribute to ${\cal P}_H$. While $\overline{p_H}$ and ${\cal P}_H$ are not trivially connected, it follows from (\ref{eq:efflong},\ref{eq:AverPop}) that optimal ${\cal P}_H$ can only be accomplished for optimal tuning of all these contributions --- what is not guaranteed by centrosymmetry for individual realisations.

\subsection{Dominant Doublet}\label{sec:DomDub}

\begin{figure}[h] 
   \centering
  	\includegraphics[width=0.49\textwidth]{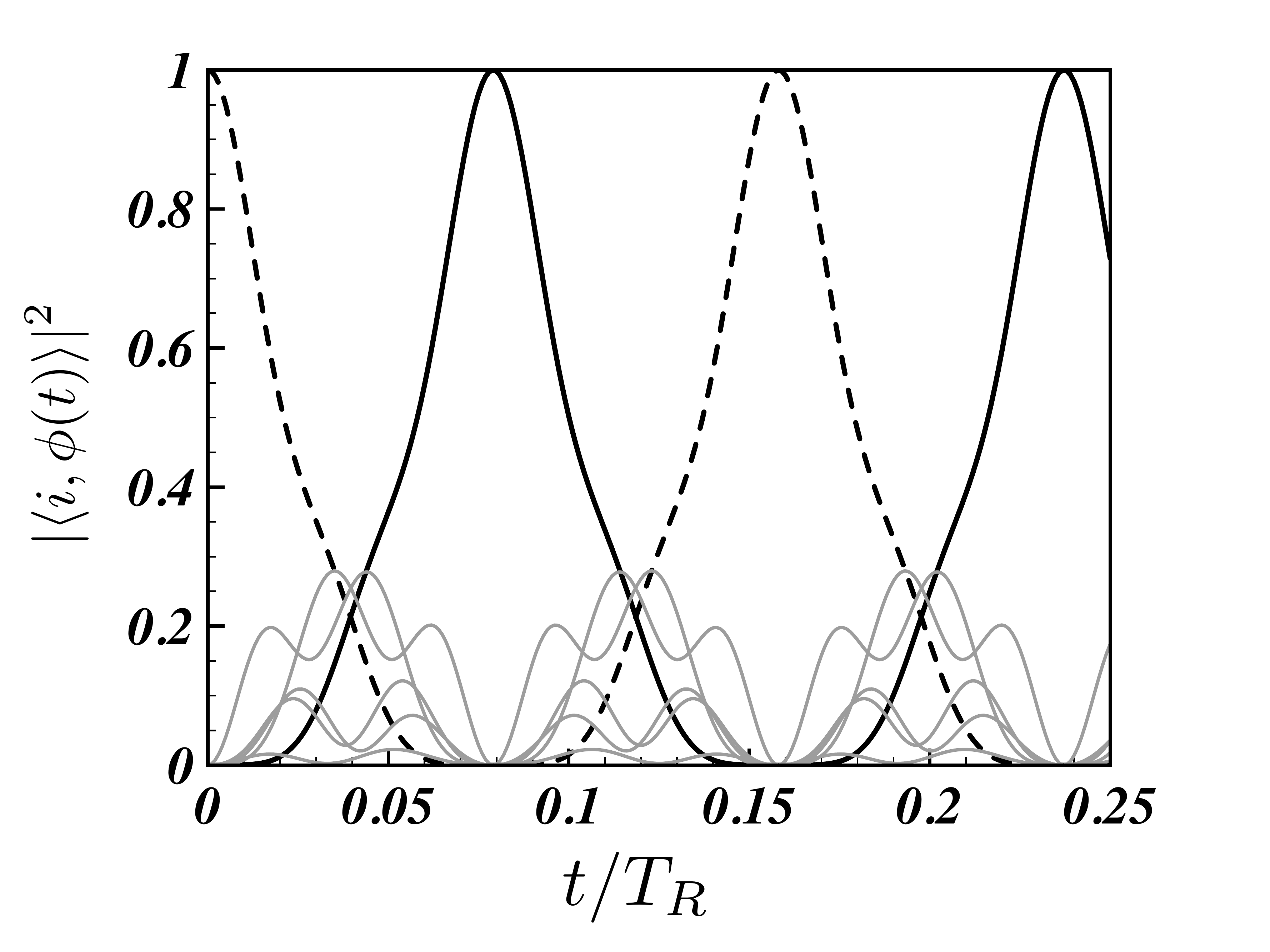}
  \caption{Population dynamics of a near-to-optimal network conformation of coupled dipoles, from \cite{Scholak:2011aa}. Mainly input (dashed black) and output (solid black) sites are populated during the dynamics, contrary to the bulk sites (gray), which exhibit weak populations never larger than approximately 30\%. Exactly this feature lies at the fundament of the dominant doublet design principle (see text). }
   \label{fig:GenAlgDynamics}
\end{figure}

We therefore need to identify an additional design principle which turns an enhanced probability of efficient transport --- as provided by centrosymmetry --- into an almost certain event. Inspection of the structures of optimal Hamiltonians generated by a genetic algorithm \cite{Scholak:2011aa} does not provide any obvious hint, but so does the time evolution of  the populations of $\ket{\rm in}$ and $\ket{\rm out}$, and of the bulk sites (see Fig \ref{fig:GenAlgDynamics}) which these Hamiltonians generate: 
Those of $\ket{\rm in}$ and $\ket{\rm out}$  are strongly indicative of the tunnelling dynamics in an effective double well potential, while the bulk sites exhibit comparably small, yet non-vanishing  populations, with the same characteristic symmetry on the time axis.slow

This observation implies that random graphs with optimal transport properties exhibit a spectral property which we have labeled {\em dominant doublet} \cite{Walschaers:2013aa}: $\ket{+}$ and $\ket{-}$, as in (\ref{eq:pm}), need to be close --- in a sense to be quantified a bit further down --- to eigenvectors $\ket{\tilde{+}}$ and $\ket{\tilde{-}}$ of $H^{+}$ and $H^-$, respectively. Under this condition, the Hamiltonian (\ref{eq:Hblock}) acquires the following, additional substructure, 
\begin{equation}\label{eq:Matrix}
H= \begin{pmatrix} E+V&\bra{\mathcal{V}^+}& &  \\ \ket{\mathcal{V}^+} & H^+_{sub}& & \\
& & E-V & \bra{\mathcal{V}^-}\\ & & \ket{\mathcal{V}^-} &H_{sub}^- \end{pmatrix},
\end{equation} 
with $\bra{\pm} H \ket{\pm} = E \pm V$, and $\ket{{\cal V}^{\pm}}$ the couplings of the states $\ket{\pm}$ to the remainder of the system. The dominant doublet assumption further implies that $\norm{{\cal V}^{\pm}}$ be sufficiently small.

Let us now exploit the dominant doublet property for a further simplification of $(\ref{eq:efflong})$. The dominant doublet's characteristic property being its dominant weight in the local density of states of the initial condition, \ie 
\begin{equation}\label{eq:domdubalpha}
\abs{\braket{\tilde{\pm}}{\pm}}^2 > \alpha \approx 1,
\end{equation}
implies that each of the two sums in (\ref{eq:efflong}) is dominated by a single term, thus
\begin{equation}\begin{split}
 \mathcal{P}_H &\approx \max_{t\in \left[0,T_R\right)}\frac{1}{4}\abs{ e^{-i t E^+} \abs{\braket{\tilde{+}}{+}}^2 -  e^{-i t E^-} \abs{\braket{\tilde{-}}{-}}^2}^2 \\
 &\gtrsim \max_{t \in [0.T_R)}  \frac{2\alpha-1}{4}\abs{ e^{-i t E^+}  - e^{-i t  E^-}}^2,
\label{eq:effdom}
\end{split}
\end{equation}
where $E^{\pm}$ in (\ref{eq:effdom}) is the eigenvalue associated with $\ket{\tilde{\pm}}$. 
The energy difference $\abs{E^+-E^-}$ of the dominant doublet states, which is reduced or enhanced with respect to the direct coupling $V$ by the collective impact of the bulk sites, now acts as an effective tunnelling rate that couples $\ket{\rm in}$ and $\ket{\rm out}$. At 
\begin{equation}
t_0=\frac{\pi}{\abs{E^+-E^-}},\label{eq:transfertime}
\end{equation} 
the transfer probability is bounded from below by $2\alpha-1$, and therefore {\em large}, since $\alpha \approx 1$. If, on top, $t_0 < T_R $, then the excitation transfer is {\em efficient} in the sense defined above. We therefore need a quantitative prediction for $\abs{E^+-E^-}$.  \\
 
Under the dominant doublet assumption perturbation theory is a valid tool to study the problem. Perturbative techniques teach us that \begin{equation}\label{eq:nondegpert}
1-\abs{\braket{\tilde{\pm}}{\pm}}^2 \approx \sum^{N/2-1}_{i=1}\frac{\abs{\braket{{\cal V}^{\pm}}{\psi_i^{\pm}}}^2}{(E\pm V -e^{\pm}_i)^2},
\end{equation}
with $\ket{\psi^{\pm}_i}$ and $e^{\pm}_i$ the eigenvectors and eigenvalues of $H^{\pm}_{sub}$, respectively. Therefore, the requirement (\ref{eq:domdubalpha}) implies a relation between $\alpha$, $\abs{\braket{{\cal V}^{\pm}}{\psi_i^{\pm}}}^2$, and $(E\pm V -e^{\pm}_i)^2$. Furthermore, $E \pm V$ each is an eigenvalue up to an energy shift $s^{\pm}$. This latter quantity can be obtained from standard perturbation theory, as
\begin{equation}\begin{split}\label{eq:Energies}
s^{\pm}= \sum_i \frac{\abs{\braket{\mathcal{V}^{\pm}}{\psi^{\pm}_i}}^2}{E\pm V - e^{\pm}_i}, \text{ such that } E^{\pm}= E \pm V + s^{\pm}. 
\end{split}
\end{equation} Notice that, for simplicity, we here present the expression that is obtained from non-degenerate perturbation theory. In the regime where $(E\pm V -e^{\pm}_i)\approx 0$, we will need to consider a more complicated expression 
(see (\ref{eq:degpert2}) in Sec.~\ref{sec:level-spacing}).\\

With $\Delta s = s^+-s^-$, it is clear that the effective tunnelling rate $\abs{E^+-E^-}$ between $\ket{\rm in}$ and $\ket{\rm out}$ can be written as 
\begin{equation}
\abs{E^+-E^-}=\abs{2V + \Delta s},\label{eq:blub}
\end{equation} 
where the direct (Rabi-like) coupling term is now ``renormalised'' by the shift $\Delta s$ imparted by the cumulative effect of the randomly placed bulk sites of the graph. Large fluctuations thereof will induce large fluctuations of the transfer efficiency. Since the statistics of $\Delta s$ is inherited from the statistics of $H^{\pm}$, we will be able to infer the statistics of the transfer efficiency, in the next chapter. 

Before doing so, let us briefly comment qualitatively on which is the implication of the dominant doublet assumption for the excitation dynamics on the random graph:
Imposing this mechanism, we greatly limited the freedom of the excitation to spread over the network, which quantum mechanically causes the typical delocalisation over the different network sites as discussed at the end of the previous subsection. As apparent from a comparison of the spectral decompositions (\ref{eq:efflong}, \ref{eq:effdom}), the eigenvectors of the Hamiltonian tell the excitation where it is allowed to go, and the dominant doublet imposes a strong incentive for the excitation to go directly from input to output (or the other way round). Yet, the time scale of the transport is set by the associated doublet {\em eigenvalues}, and these may be strongly affected by the remainder of the spectrum, via (\ref{eq:Energies}), as we will see hereafter.

\section{Statistics of Transfer Time Scales}\label{sec:Statistics}

We have so far reformulated our initial transport problem in terms of a spectral doublet structure which is amended by the perturbative coupling to some bulk states described by random matrices. This is a general scenario which is well-known under the name {\em chaos assisted tunnelling} (CAT) \cite{PhysRevE.50.145} in the area of quantum chaos \cite{Madronero:2006aa}, and also reminiscent of transport problems in mesoscopic physics \cite{ImryBook}. The fundamental idea is that the dynamical and/or transport properties in some predefined degree of freedom can be {\em dramatically} modified by the nonlinear coupling to some other degrees of freedom, incarnated, \eg, by a classical driving field \cite{Buchleitner:1998aa,Schlagheck:2003aa,Wimberger:2001aa,PhysRevE.57.1458}, or by further coordinates of configuration space \cite{PhysRevLett.95.263601,Madronero:2005aa,PhysRevE.50.145}. In the specific context of photosynthetic light harvesting, ideal candidates for such additional degrees of freedom are provided by those of the protein scaffold, which fix the boundary conditions for the electronic dynamics and excitations \cite{Falke:2014aa,Hildner:2013aa,Mancal:2012aa}. If these additional degrees of freedom themselves exhibit sufficiently complex dynamics, their coupling to the transporting degree of freedom will induce strong fluctuations in the transport properties of interest. We now import the random matrix theory (RMT) of CAT to derive analytical predictions for the statistics of the transfer efficiencies (\ref{eq:effdom}) and 
times (\ref{eq:transfertime}), and in particular discuss the necessary amendments  of the available theory to match the details of our model.

\subsection{How to Obtain the Distribution of Transfer Times}\label{sec:ObtDist}

 The distribution of $s^{\pm}$ is already known in terms of CAT, with $E,V=0$, and we will therefore strongly rely on the results of \cite{Leyvraz:1996aa,PhysRevE.57.1458}. Note, however, that already \cite{Lopez:1981aa} argues under very general assumptions that the distribution of this type of quantity should always be a Cauchy distribution, irrespective of whether the $e^{\pm}_i$ strictly derive from GOE or from some other type of random Hamiltonian. This is important in our present context, since the biological functional units which inspire the present study are unlikely to realise GOE statistics in the strict sense. Moreover, \cite{Leyvraz:1996aa,PhysRevE.57.1458} provide us with clear insight in the parameters determining the Cauchy distribution, for a setup which is close to ours. Adopting the mathematical language of \cite{Leyvraz:1996aa,PhysRevE.57.1458}, we obtain that, when $E=V=0$, the distribution of $s^{\pm}$ is given by
\begin{equation}\label{eq:cauchylit}\begin{split}
P(s^{\pm})&= \frac{1}{\pi}\frac{\sigma^{\pm}}{\left(\sigma^{\pm} \right)^2+(s^{\pm}-s_0^\pm)^2}=\text{Cauchy}(s_0^\pm,\sigma^\pm),\\ &\text{with } \sigma^{\pm}=\pi \frac{ \overline{\abs{\braket{{\cal V}^{\pm}}{\psi_i^{\pm}}}^2}}{ \Delta}\ ,\ s_0^\pm=0\ ,
\end{split}
\end{equation}
where we assume that $\overline{\abs{\braket{{\cal V}^{\pm}}{\psi_i^{\pm}}}^2}=\overline{\norm{\mathcal{V^{\pm}}}^2}(N/2-1)^{-1/2},$ with  $\overline{\norm{\mathcal{V^{\pm}}}^2}$ a measure for the average interaction strength between $\ket{ \pm}$ (and, thus, also $\ket{{\rm in}}$ and $\ket{{\rm out}}$) and the bulk states. The parameter $\Delta$ expresses the mean level spacing in the vicinity of $0$ \cite{Leyvraz:1996aa,PhysRevE.57.1458}.

In contrast to $E=V=0$ in \cite{Leyvraz:1996aa,PhysRevE.57.1458}, we need to accommodate for $E\pm V \neq 0$. This can be accomplished using the results of \cite{PhysRevE.68.046124} and realising that the curvatures presented in eq. (5) of \cite{PhysRevE.68.046124} are closely related to the energy shifts. Indeed, the shifts' distribution is given by eq. (49) of \cite{PhysRevE.68.046124}, with $\sigma_2^2$, $\sigma^2$ and $\lambda$ in \cite{PhysRevE.68.046124} substituted by  $\overline{\norm{\mathcal{V^{\pm}}}^2}$, $2\xi^2$, and $E\pm V$, respectively, in our present nomenclature. 

Note that, in \cite{PhysRevE.68.046124}, $\rho(\lambda)$ is the density of states, given by Wigner's semicircle \cite{BohigasLesHouches1989} in the GOE. In the standard GOE scenario, the mean level spacing is known to only change slightly throughout the bulk of the spectrum, and it can be estimated by the radius of the semicircle \cite{BohigasLesHouches1989}. In our present context, this would imply that we can use $\xi$ as a parameter just as well as $\Delta$. Note, however, that the matrices $H_{sub}^{\pm}$ of our model are in general {\em not} GOE matrices, since they are obtained by post-selection of that matrix sub-ensemble of structure (\ref{eq:Hblock}) which exhibits a dominant doublet as defined by (\ref{eq:domdubalpha}). This means that the post-selected ensemble does not obey Wigner-Dyson statistics, and thus $\rho(\lambda)$ is typically {\em not} the semicircle distribution. As essential consequence, the relation between the radius of the semicircle and the local mean level spacing no longer holds; we can no longer relate the global quantity $\xi$ to the local parameter $\Delta$! Moreover, it turns out, as extensively discussed in Section \ref{sec:level-spacing} below, that $\Delta$ can vary {\em strongly} throughout the spectrum. In our derivation, the relevant quantity is the mean level spacing {\em in the vicinity} of $E \pm V$, which we will refer to as $\Delta_{\rm loc}.$

Given (\ref{eq:transfertime},\ref{eq:blub}), we need to infer  the distribution of $\Delta s=s^+-s^-$. To do so, we can use simple properties of the Cauchy distribution. The fact that $s^+ \sim \text{Cauchy}(s^{+}_0,\sigma^+)$ and $s^- \sim \text{Cauchy}(s_0^-,\sigma^-)$ implies that $s^+-s^- \sim \text{Cauchy}( s^+_0-s_0^-, \sigma^+ + \sigma^-)$, which follows from the Cauchy distribution being a {\em stable} distribution \cite{Fama:1968aa}. In order to simplify notation, we define $s_0= s^+_0-s_0^-$ and $\sigma=\sigma^+ + \sigma^-$, to obtain: \begin{equation}\begin{split}&P(\Delta s)= \frac{1}{\pi}\frac{\sigma}{\sigma^2+(\Delta s-s_0)^2},\\&\text{with} \quad s_0= 2 V \frac{\overline{\norm{\mathcal{V^{\pm}}}^2}}{2 \xi^2},\\
&\text{and} \quad \sigma = 2\pi \frac{\overline{\norm{\mathcal{V^{\pm}}}^2}}{(N/2-1)\Delta_{\rm loc}}.\end{split}\end{equation}
where we used that $\overline{\norm{\mathcal{V}^+}^2}=\overline{\norm{\mathcal{V}^-}^2}=\overline{\norm{\mathcal{V}}^2}.$ This follows from $\norm{\mathcal{V}^+}^2$ and $\norm{\mathcal{V}^-}^2$ being independent stochastic variables which are identically distributed, a property which they inherit from $H^+$ and $H^-$ being independent and identically distributed, and hence have the same expectation value. 

The distribution of $\Delta s$ is but a first step to derive the distribution of $T_R/t$. The expressions for $t_0$ and $T_R$, using (\ref{eq:transfertime},\ref{eq:blub}), imply that \begin{equation}\label{eq:tTR}
\frac{T_R}{t}=\abs{1-\frac{\Delta s}{2V}}.
\end{equation}
Since $E$ and $V$ are still considered to be fixed, we again use that the Cauchy distribution is stable \cite{Fama:1968aa}: This implies that, if $\Delta s \sim  \text{Cauchy}\left(s_0, \sigma\right),$ then \begin{equation}1-\frac{\Delta s}{2V} \sim \text{Cauchy}\left(1-\frac{s_0}{2V}, \frac{\sigma}{2V}\right).\end{equation} The distribution of the absolute value  $\abs{1-\frac{\Delta s}{2V}}$ thus reads: 
\begin{equation}\label{eq:DistFixed}\begin{split} 
P\left(\abs{1-\frac{\Delta s}{2V}}=x\right)&=\frac{1}{\pi}\left(\frac{\gamma}{{\gamma}^2+(1+x_0+x)^2 }+\frac{\gamma}{{\gamma}^2+(1+x_0-x)^2 }\right),\\
&\text{with} \quad x_0 = \frac{\overline{\norm{\mathcal{V^{\pm}}}^2}}{2 \xi^2},\\
&\text{and} \quad \gamma = \frac{1}{V}\frac{\pi \overline{\norm{\mathcal{V}}^2}}{(N/2-1) \Delta_{\rm loc}}.
\end{split}
\end{equation}

We finally need to account for the fact that $E$ and $V$ are themselves stochastic variables, and we therefore need to average over their respective distributions. However, as shown in Section \ref{sec:LaplaceMethod} below, the probability distribution of $V$ is strongly peaked and, therefore, dominated by its mean value $\overline{V}$. Given this dominant behaviour of the mean, it is usually a reasonable approximation to replace $V$ by $\overline{V}$ rather than exactly performing the integration. This approximation is what is called an {\em annealed approximation} \cite{PhysRevA.45.6056}, and leads to
\begin{equation}\label{eq:18}
\gamma \approx \frac{1}{ \overline{V}}\frac{ \overline{\norm{\mathcal{V}}^2}}{(N/2-1) \Delta_{\rm loc}},
\end{equation} 
where $\Delta_{\rm loc}$ --- the local mean level-spacing of energy levels in the vicinity of the energy $E \pm V$--- and the value $\overline{V}$ still are to be determined. 

Since the dominant doublet constraint modifies the local properties of the $H^{\pm}_{sub}$ ensemble around $E \pm V$, we cannot simply import the results available for GOE. We will therefore present a derivation of $\Delta_{\rm loc}$ in Section \ref{sec:level-spacing} hereafter, and already warn the reader that this section will be rather technical and not extremely elegant, however with the useful result \begin{equation}\label{eq:rhoMod_App}\Delta_{\rm loc} \approx \frac{2 \pi \xi}{\sqrt{N/2-1}}.\end{equation} Section \ref{sec:LaplaceMethod} below will provide the derivation of the parameter $\overline{V}$, which is mainly based on a Laplace approximation for the integration, and yields \begin{equation}\label{eq:vbarFinal}
\overline{V}\approx \frac{2 \pi \xi}{e N \sqrt{N/2-1}}. \end{equation}\\

With the explicit expressions (\ref{eq:rhoMod_App}) and (\ref{eq:vbarFinal}) in (\ref{eq:18}), we ultimately obtain from (\ref{eq:DistFixed}):
\begin{equation}
\label{eq:Dist}\begin{split} 
&P\left(\frac{T_R}{t}=x\right)=\frac{1}{\pi}\left(\frac{s_0}{{s_0}^2+(1+x_0+x)^2 }+\frac{s_0}{{s_0}^2+(1+x_0-x)^2 }\right)\, ,\\
& \quad\text{with} \quad s_0=\frac{\overline{\norm{\mathcal{V}}^2} N e }{4 \pi \xi^2},\,\\
& \quad \text{and} \quad x_0=\frac{\overline{\norm{\mathcal{V}}^2}}{2\xi^2}.
\end{split}
\end{equation}
This is our final result for the distribution of the excitation transfer times generated by centrosymmetric Hamiltonians of the form (\ref{eq:Matrix}) with dominant doublet strength $\alpha$. The relationship between $\alpha$, which is not 
explicit in (\ref{eq:Dist}), and $\overline{\norm{\mathcal{V}}^2}$ will be derived in Sec.~\ref{sec:level-spacing} below, see (\ref{eq:relAlpha}).

\subsection{Scaling Properties of Characteristic Transfer Times}

From the thus obtained Cauchy distribution for $T_R/t$ we can obtain a good understanding of the probability of finding ${\cal P}_H$ close to one. According to (\ref{eq:effdom}), it is clear that $\mathcal{P}_H\geqslant 2\alpha-1$ close to one if  $t=\pi/\abs{2V+\Delta s}<T_R$.  Therefore, we can infer the probability that $T_R/t$ is larger than one by straightforward integration over the corresponding range in (\ref{eq:Dist}). The result reads:
 \begin{equation}\begin{split}\label{eq:ProbLargerAlpha2}
P\left(\frac{T_R}{t} > 1\right) = 1-\frac{1}{\pi} \arctan \left( \frac{4 \pi \xi^2}{\overline{\norm{\mathcal{V}}^2} N e}\left(1-\frac{\overline{\norm{\mathcal{V}}^2}}{2\xi^2}\right)\right).
\end{split}
\end{equation}
It follows that the probability for fast and efficient transport {\em increases} with the size $N$ of the network. As $N$ grows very large, we obtain
\begin{equation}\label{eq:Asympt1}
P\left(\frac{T_R}{t} > 1\right) \approx 1- \frac{4 \xi^2}{\overline{\norm{\mathcal{V}}^2} N e}.
\end{equation}
In other words, the tail of the distribution in eq. (\ref{eq:Dist}) grows heavier with increasing $N$ and therefore more and more realisations enhance the transport. The origin of this scaling can be traced back to the direct (in-out) coupling $V$, since $N$ enters through $\overline{V}$. The coupling is the smallest number in absolute value of a set of $N/2$ normally distributed variables, and, as explained in Section \ref{sec:LaplaceMethod} below, for a fixed density of states its expectation value decreases $\propto N^{-3/2},$ in leading order. In large systems, the direct tunnelling from input to output will be negligible, and the intermediate sites provide a considerable boost to the transport (much in the spirit of CAT \cite{PhysRevE.50.145}). Thus, if we compare the time scale of the direct coupling, $T_R$, to the effective transport time $t$, we should find $t < T_R$ with high probability. This intuition perfectly matches the result 
displayed in Fig.~\ref{fig:time} below.

Alternatively, when studying systems where the direct coupling is fixed to a value $V^*$ for all realisations of the networks' conformation, a very different scaling is obtained (by suitable integration of eq.~(\ref{eq:DistFixed}) -- rather than of (\ref{eq:Dist}), due to the explicit dependence on $V$ in (\ref{eq:DistFixed})):  \begin{equation}\begin{split}\label{eq:ProbLargerAlphaFixed}
P\left(\frac{T_R}{t} > 1\right) = 1-\frac{1}{\pi} \arctan \left(\frac{2 V^* \xi \sqrt{N/2-1}}{\overline{\norm{\mathcal{V}}^2}}\left(1-\frac{\overline{\norm{\mathcal{V}}^2}}{2\xi^2}\right)\right).
\end{split}
\end{equation}
Now we find that, in the limit of large $N$, this expression scales as 
\begin{equation}\label{eq:Asympt2}
P\left(\frac{T_R}{t} > 1\right) \approx  \frac{1}{2} + \frac{\overline{\norm{\mathcal{V}}^2}}{\pi V^* \xi \sqrt{2 N}},\end{equation}
\ie~the relative weight of conformations which enhance the transport {\em decreases} with $N$ --- though remains bounded from below by $50\%$. Since, in this regime, the direct tunnelling from $\ket{{\rm in}}$ to $\ket{{\rm out}}$ always has the same strength, we can thus conclude that increasing the system size in this post-selected ensemble has a negative impact on the chaos-assisted tunnelling contribution to the transport ---  the peak around $T_R/t=1$ in the Cauchy distribution (\ref{eq:DistFixed}), is enhanced at the expense of the tail.

The two asymptotic scaling laws (\ref{eq:Asympt1}, \ref{eq:Asympt2}) can be given a more physical interpretation: If, as in the molecular networks at the heart of photosynthetic light-harvesting (which inspired our model), coupling strength is synonymous to spatial separation, then increasing $N$ {\em at fixed spatial density}, and thus literally increasing the spatial size of the network, leads to (\ref{eq:Asympt1}). Alternatively, keeping the spatial size of the network fixed and increasing the packing density by increasing $N$ leads to (\ref{eq:Asympt2}).

\begin{figure}[t] 
   \centering
  	\includegraphics[width=0.49 \textwidth]{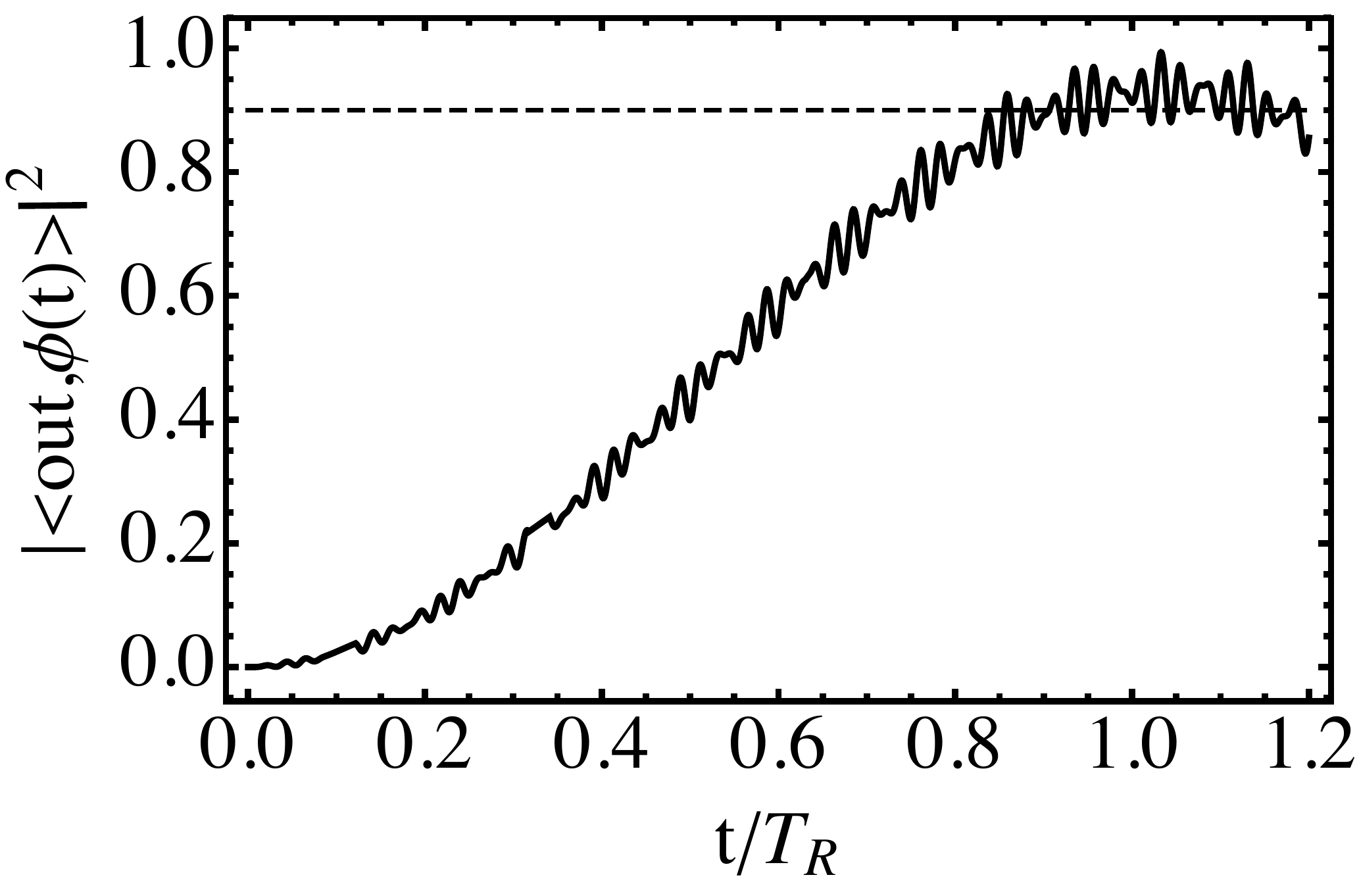}
  \caption{Probability $\abs{\braket{\text{out}}{\phi(t)}}^2$ to find the excitation at the output site, for a single realisation of the network Hamiltonian (\ref{eq:Matrix}). The value $2\alpha-1$ is indicated by a dashed line. Even though the transfer time $t=1.03 T_R$ for this realisation, ${\cal P}_H > 2\alpha-1$.}
   \label{fig:Dynamics}
\end{figure}

In closing this part of our discussion, let us also emphasise that the probability given in (\ref{eq:ProbLargerAlpha2}) is only a lower bound of the probability to obtain ${\cal P}_H > 2\alpha-1$. In order to understand this, let us reconsider equation (\ref{eq:effdom}): The time $t=\pi/\abs{2V+\Delta s}$ is the point in time when $\abs{ \exp({-i t E^+})  - \exp({-i t  E^-})}^2/4$ reaches its largest possible value. Nevertheless, for a specific realisation of the disorder, we may find other (and in particular earlier) moments in time at which already $\abs{ \exp({-i t E^+})  - \exp({-i t  E^-})}^2/4>2\alpha-1$. These realisations are not included in  (\ref{eq:ProbLargerAlpha2}) (which was derived by using the relation (\ref{eq:transfertime},\ref{eq:blub})), although ${\cal P}_H> 2\alpha-1$. As an example, Fig.~\ref{fig:Dynamics} shows a realisation of the time dependence of the output population for which $T_R/t = 0.970874$, and indicates the value $2\alpha-1$ by a dashed line. Since $t > T_R$, we do not account for this in our estimate (\ref{eq:ProbLargerAlpha2}) of efficient realisations, even though it clearly exhibits ${\cal P}_H > 2\alpha-1$.\\

As a final remark of this section, and as an important intermediate result, let us emphasise that the desired transport properties of the network as described above do {\em not} depend on the details of the individual networks' structures. 
Indeed, only course grained and somewhat easily controllable quantities --- the spectral density $\xi$ of the bulk states, and the average coupling strength $\overline{\norm{\mathcal{V}}^2}$ of the input and output site to the bulk --- fully determine the distribution (\ref{eq:Dist}).

\subsection{The Mean Level Spacing $\Delta_{\rm loc}$ in the Vicinity of $E\pm V$}
\label{sec:level-spacing}

\begin{figure}[h] 
   \centering
  	\includegraphics[width=0.49 \textwidth]{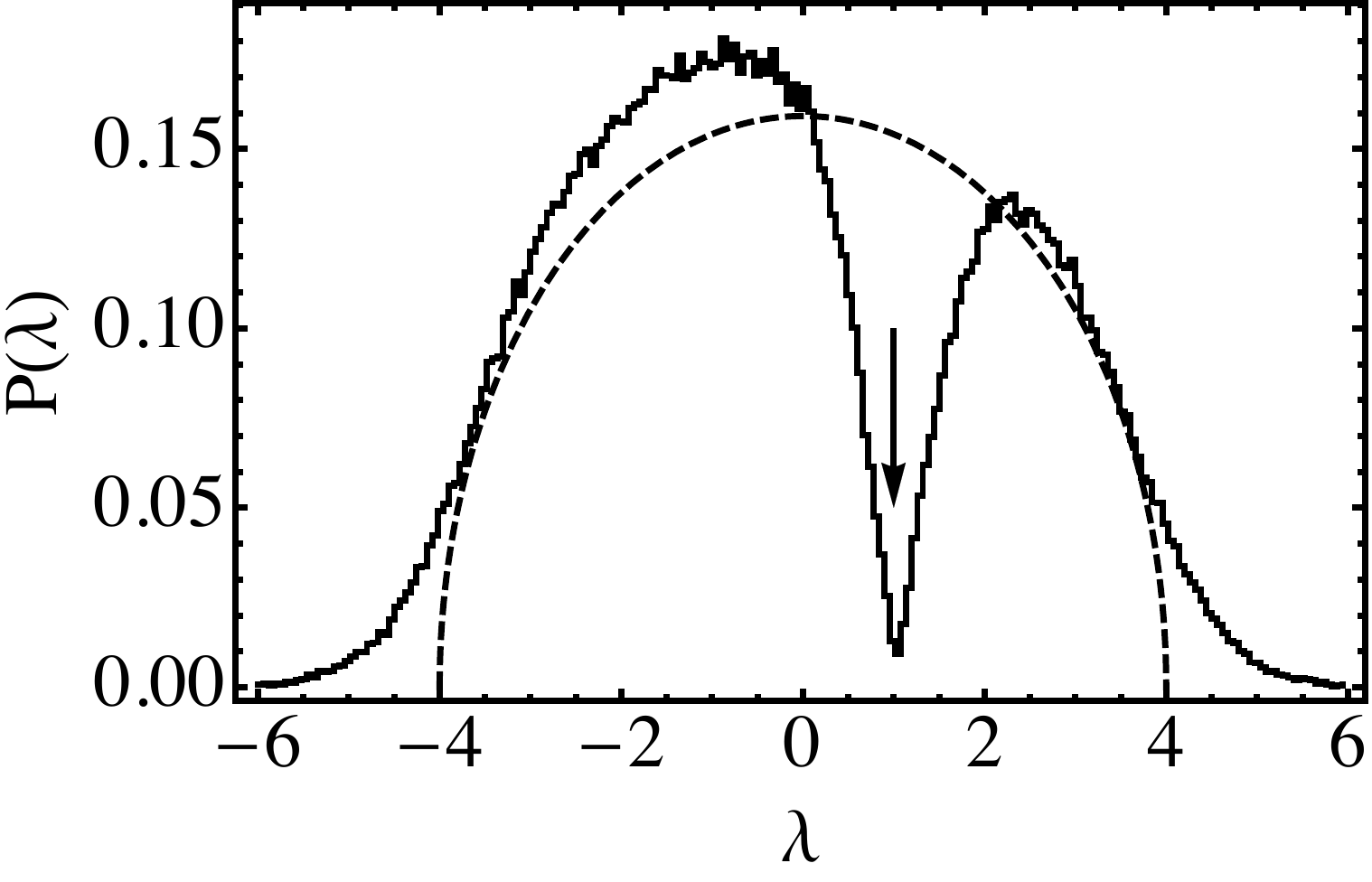}
  \caption{Density of states of $H^+_{sub},$ for $N=10$ and $\xi=2$, with fixed $E+V=1$ (arrow) to highlight the effect of a dominant doublet in the vicinity of this energy level. In contrast to the Wigner semicircle (dashed line), valid for the GOE ensemble with $N \rightarrow \infty$, the density of states exhibits a cusp at $\lambda = E
  +V$.}
   \label{fig:DOS}
\end{figure}

Now that a global picture has been established, we need to understand the technical details required to obtain an expression for $\Delta_{\rm loc}$, the mean level-spacing near the energy $E \pm V$ which entered (\ref{eq:Dist}) through (\ref{eq:DistFixed}, \ref{eq:18}). It was already indicated in Section \ref{sec:model} that the dominant doublet constraint is somewhat more subtle than the mechanism of chaos assisted tunnelling, where this mean level-spacing is known {\em a priori}. The dominant doublet in our model can be seen as a strong demand of eigenvector localisation (\ref{eq:domdubalpha}). Since, in our present 
work, we sample centrosymmetric Hamiltonians and {\em post-select} realisations where a dominant doublet is present, a strong modification of the local mean level spacing around the energy $E\pm V$ can be induced.
This effect is also apparent from the density of states, shown in Figure \ref{fig:DOS}: Wigner's semicircle 
-- to be expected from RMT \cite{BohigasLesHouches1989} -- is garnished by a cusp, centred around $E+V$(in the figure fixed at  $E+V=1$). The 
key approach to deriving an estimate for $\Delta_{\rm loc}$ is the assumption that it is essentially the same quantity as the width of the cusp, which we now set out to determine.\\

To reach a quantitative understanding of the cusp effect, we must be able to treat the eigenvalues $e^{\pm}_i$ of $H^{\pm}$ which are close to $E \pm V$. Here, we find strong repulsion between the energy levels, causing the cusp. For an exact description of the effect, we must include the possibility of (near-)degeneracy between $E \pm V$ and one of the $e^{\pm}_i$, and thus use degenerate perturbation theory. Therefore, we first consider the degenerate variant of (\ref{eq:nondegpert}):
\begin{equation}\label{eq:degpert2}
1-\abs{\braket{\tilde{\pm}}{\pm}}^2 \approx \frac{1}{2} \sum^{N/2-1}_{i=1} \left(1-\left[1+4\frac{\abs{\braket{{\cal V}^{\pm}}{\psi_i^{\pm}}}^2}{(E\pm V -e^{\pm}_i)^2}\right]^{-1/2}\right).
\end{equation}
All special effects caused by the dominant doublet originate from this expression, via (\ref{eq:domdubalpha}). The requirement that the left hand side of (\ref{eq:degpert2}) be smaller than $1-\alpha$ imposes constraints on the possible values which \begin{equation}\label{eq:D}D:=\min_i\abs{E\pm V -e^{\pm}_i}\end{equation} can take. As the quantity $D$ is directly related to the cusp in Fig. \ref{fig:DOS}, it will form the cornerstone to our estimate of $\Delta_{\rm loc}$.

First, we observe that there are three parameters in (\ref{eq:degpert2}) which must be controlled to fulfil the dominant doublet constraint: $D$, $\norm{\mathcal{V}}^2$, and $\alpha$. Of these, only the last one is controlled directly in our setup. Looking at the right hand side of (\ref{eq:degpert2}), one sees that the dominant doublet regime is reached for $\norm{\mathcal{V}}/D$ sufficiently small, such that this right hand side of the equation vanishes and $1-\abs{\braket{\tilde{\pm}}{\pm}}^2$ is close to zero. While $\norm{\mathcal{V}}^2$ can be measured rather straightforwardly in our simulations, we require an estimate of $D$ in terms of the other known parameters: $\alpha$, $\xi$ and $\overline{\norm{\mathcal{V}}^2}$. 

To obtain such an estimate, we focus on two limiting cases, which we both expect to encounter in the same ensemble of post-selected Hamiltonians. Moreover, each of these cases will impose constraints on the possible range of the parameters $\norm{\mathcal{V}}$ and $D$. As mentioned, the dominant doublet implies that $\norm{\mathcal{V}}/D$ should be small, what implies that $\norm{\mathcal{V}}$ is sufficiently small, or that $D$ is sufficiently large. The two limiting cases exactly boil down to these scenarios: In the first limiting case, we will consider Hamiltonians where all eigenvalues $e^{\pm}_i$ are outside of the cusp region of Fig.~\ref{fig:DOS}. In this regime, the dominant doublet imposes constraints on $\norm{\mathcal{V}}$. In the other limiting case, we investigate what happens when one of the $e^{\pm}_i$ lingers inside the cusp region of Fig.~\ref{fig:DOS}, which leads to constraints on $D$. Throughout these calculations, even though mathematically somewhat unsound, we assume that $\norm{\mathcal{V}}$ and $D$ are two independent statistical quantities. Finally, once the two limiting scenarios have been considered, we combine the two constraints, as they should both hold for the complete ensemble, and formulate an estimate for the width of the cusp.

{\bf The first limiting case} is given by network realisations where all eigenvalues $e_i^{\pm}$ exhibit a considerable distance from $E \pm V$, far away from the observed cusp in Fig. \ref{fig:DOS}. Therefore, all terms in the sum (\ref{eq:degpert2}) contribute equally. This leads to the approximation that the expectation value of a single one of these terms is $(1-\alpha)/(N/2-1)$.  Rather than (\ref{eq:degpert2}), we can then use (\ref{eq:nondegpert}), \ie  \begin{equation}
1-\abs{\braket{\tilde{\pm}}{\pm}}^2 \approx \sum^{N/2-1}_{i=1}\frac{\abs{\braket{{\cal V}^{\pm}}{\psi_i^{\pm}}}^2}{(E\pm V -e^{\pm}_i)^2}.
\end{equation} 
On the level of averages, the dominant doublet condition tells us thus that \begin{equation}\label{eq:ThatOne}
\overline{\frac{1-\alpha}{N/2-1}} \approx  \overline{\left(\frac{\abs{\braket{{\cal V}^{\pm}}{\psi_i^{\pm}}}^2}{(E\pm V -e^{\pm}_i)^2}\right)}.
\end{equation}
Here we assume that, since the $e^{\pm}_i$ stay far away from the cusp, and therefore do not feel the ``repulsion'' from $E\pm V$, $E$ and $V$ can be approximately treated as independent variables. The variance of $V$ --- as its statistics is described by extreme value theory \cite{deHaan2007extreme}, 
see Section \ref{sec:LaplaceMethod} below --- is neglected as the distribution of $V$ is strongly peaked around $\overline{V}$. Furthermore, we approximate the distribution of the $e_i$ (locally) by a semicircle law. The crude approximation that each term in (\ref{eq:nondegpert}) provides a similar contribution leads to
\begin{equation}\label{eq:ThisOne}
\sqrt{\frac{1-\abs{\braket{\tilde{\pm}}{\pm}}^2}{N/2-1}} \approx \frac{\abs{\braket{{\cal V}^{\pm}}{\psi_i^{\pm}}}}{\abs{E\pm V -e^{\pm}_i}}, \text{ for all } i.
\end{equation} 
Comparing (\ref{eq:ThatOne}) to (\ref{eq:ThisOne}), we get $\overline{1-\abs{\braket{\tilde{\pm}}{\pm}}^2} \approx 1-\alpha$. If we now assume that $\braket{{\cal V}^{\pm}}{\psi_i^{\pm}}$ is normally distributed, with zero mean and variance $\overline{\norm{\mathcal{V}}^2}/(N/2-1)$, and that the $e_i$ obey a semicircle law  --- as one expects from RMT \cite{Leyvraz:1996aa} ---, we find 
\begin{equation}\label{eq:Const1}
\sqrt{1-\alpha} \approx \sqrt{\frac{2 \overline{\norm{\mathcal{V}}^2}}{\pi \xi^2}}.
\end{equation}

\begin{figure}[t] 
   \centering
  	\includegraphics[width=0.5\textwidth]{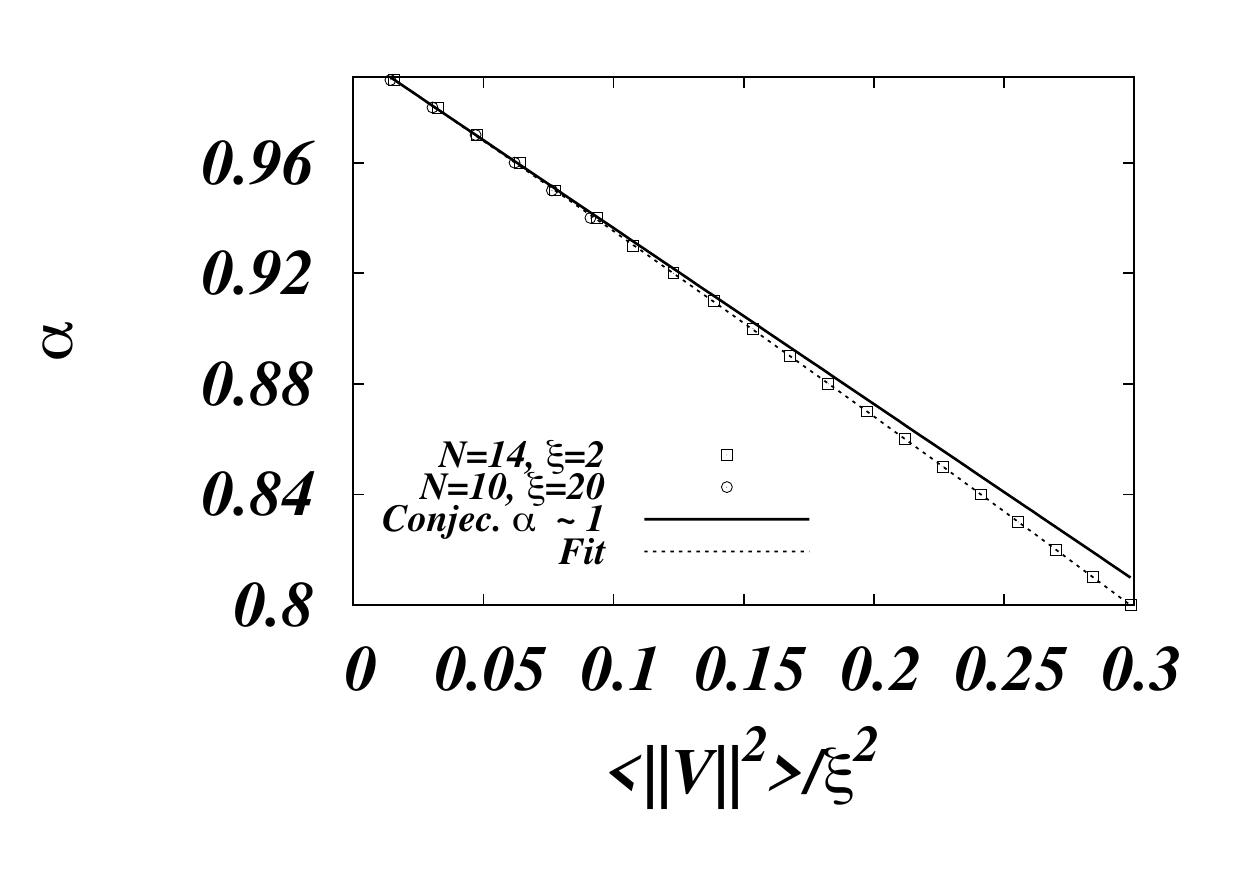}
  \caption{Dependence of $\alpha$ on $\overline{\norm{\mathcal{V}}^2}/\xi^2,$ for different network realisations. In order to extract the constant $C$ in (\ref{eq:LinFit},\ref{eq:fit}), a fit is performed. The conjectured curve for $\alpha \approx 1$, where $C = 2/\pi$, is given by the solid line.}
   \label{fig:Fit}
\end{figure}

We validate this result by numerical data (see Section \ref{sec:sim}) and approximate \begin{equation}\label{eq:LinFit}\alpha \approx 1 - C \frac{\overline{\norm{\mathcal{V}}^2}}{\xi^2},\end{equation} with $C$ as a fit parameter. The numerical dataset is obtained  by scanning $\alpha$ from $0.99$ to $0.8$, for fixed $\xi=2$ and $N=14$. For each value of $\alpha$ we extract $\overline{\norm{\mathcal{V}}^2}$. We also inspected data with $\alpha \in [0.94, 0.99]$, $\xi=20$, and $N=10$. Figure \ref{fig:Fit} suggests a linear dependence as in (\ref{eq:LinFit}). However, since the ansatz (\ref{eq:LinFit}) results from perturbation theory, it appears reasonable to add a  term quadratic in $\overline{\norm{\mathcal{V}}^2}/\xi^2$, for $\alpha \approx 0.8$. We thus fit the data to the form \begin{equation}\label{eq:fit}
\alpha \approx 1 - C \frac{\overline{\norm{\mathcal{V}}^2}}{\xi^2} - b \left(\frac{\overline{\norm{\mathcal{V}}^2}}{\xi^2}\right)^2,
\end{equation}
and obtain the following result:
\begin{center} 
\begin{tabular}{c|ccccc} 
\text{} & \text{Estimate} & & \text{Standard Error} \\ 
\hline 
C & 0.636789 & & 0.00218418 \\ 
b & 0.111501 & & 0.00933118\\ 
\end{tabular} 
\end{center} 
 By definition (\ref{eq:domdubalpha}), the dominant doublet is found where $\alpha \approx 1$ and thus $\overline{\norm{\mathcal{V}}^2}/\xi^2$ is small. Therefore we can finally ignore the second order term in equation (\ref{eq:fit}) and obtain that $\alpha \approx 1-0.636789 \times\overline{\norm{\mathcal{V}}^2}/\xi^2$. As the estimate $C \approx 2/\pi$ falls into the error margin of our numerically generated data, we conclude that
\begin{equation}\label{eq:relAlpha}
1-\alpha \approx \frac{2 \overline{\norm{\mathcal{V}}^2}}{\pi \xi^2}.
\end{equation}

Since this limiting case was defined above as the one where the eigenvalues $e^{\pm}_i$ are far away from the cusp, and thus do not experience the repulsion which must be felt as they approach $E \pm V$ (this exactly causes the cusp seen in Figure \ref{fig:DOS}), we will consider a second limiting scenario in order to probe the smallest possible value of $D$ as given by (\ref{eq:D}).\\

{\bf The second limiting case} is when one eigenvalue $e_i^{\pm}$ approaches  $E \pm V$ at a minimum distance $D_{\rm min}$ (where the minimum is taken over the entire ensemble) such that we find the dominant doublet with probability one, with $\braket{\psi^{\pm}_i}{\cal V^{\pm}}$ still a normally distributed stochastic variable. This implies that the sum in (\ref{eq:degpert2}) be dominated by a single term. Ultimately our goal is to determine $D_{\rm min}$, and to do so we study the statistics of a single term $\tau$ in (\ref{eq:degpert2}), leaving the resonance denominator, (\ref{eq:D}), as a free parameter. This term is a stochastic quantity, and we can obtain its probability density as 
\begin{equation}
P_D(\tau)=\int_{\mathbb{R}} {\rm d}v N(v) \delta\left(\tau - \frac{1}{2}\left[1-\left(1+4\frac{v^2}{D^2}\right)^{-1/2}\right]\right).
\end{equation}
\noindent  $v=\braket{{\cal V}^{\pm}}{\psi_i^{\pm}}$ is again normally distributed, with zero mean and variance $\overline{\norm{\mathcal{V}}^2}/(N/2-1),$ and we denote the Gaussian probability density function by $N(v)$.

The integration can be performed straightforwardly using properties of the Dirac delta function. As the dominant doublet arises in a regime where $1-\alpha \approx 0$, we obtain from (\ref{eq:degpert2}) that also $\tau$ must me close to zero, hence we can focus on the leading scaling behaviour in $\tau \rightarrow 0$, from which we obtain
\begin{equation}\label{eq:DistTerm}
P_D(\tau) \approx \frac{D \sqrt{N-2}}{4 \sqrt{\pi\overline{ \norm{\mathcal{V}}^2} \tau}}.
\end{equation} 
Remember that the dominant doublet was imposed as a strict constraint (\ref{eq:domdubalpha}), which implies that  $\abs{\braket{\tilde{\pm}}{\pm}}^2$ is {\em always} larger than $\alpha$. Since we are now studying the case where one term $\tau$ dominates the sum in (\ref{eq:degpert2}), $\tau$ must {\em always} be smaller than $1-\alpha$. This condition translates to
\begin{equation}
{\rm Prob} (\tau \leqslant 1- \alpha)=\int_0^{(1-\alpha)}{\rm d}t P_D(\tau)=1,
\end{equation} 
and, with (\ref{eq:DistTerm}), defines an equation which can be solved to obtain the smallest possible value for $D$, which is denoted by $D_{\rm min}$: \begin{equation}\label{eq:Const2}
D_{\rm min}=\frac{    \sqrt{2 \pi \overline{ \norm{\mathcal{V}}^2}}}{ \sqrt{(1-\alpha)(N/2-1)}}.
\end{equation}
 $D_{\rm min}$ gives the closest allowed distance between $E\pm V$ and an eigenvalue $e_i$ of $H^{\pm}_{sub}$ to ensure $\abs{\braket{\tilde{\pm}}{\pm}}^2>\alpha$. \\

{\bf Combining the constraints} (\ref{eq:Const1}) and (\ref{eq:Const2}), which connect the parameters $\alpha$, $\overline{\norm{{\cal V}}^2}$, $\xi$ and $D_{\rm min}$, we can express $D_{\rm min}$ as: \begin{equation}\label{eq:D2}
D_{\rm min}=\ \frac{ \pi \xi}{ \sqrt{N/2-1}}.
\end{equation}
With (\ref{eq:D2}), we obtain the following strong {\em conjecture} for $\Delta_{\rm loc}$: 
\begin{equation*}\tag{\ref{eq:rhoMod_App}}\Delta_{\rm loc} \approx \frac{2 \pi \xi}{ \sqrt{N/2-1}},\end{equation*}
where we used that $D_{\rm min}$ is the minimal distance between $E\pm V$ and an eigenvalue $e^{\pm}_i$ which can establish a dominant doublet. Since we are interested in the distance between two eigenvalues $e^{\pm}_i$ and $e^{\pm}_j$, which we approximate by the width of the cusp in Fig.~\ref{fig:DOS}, we acquire an extra factor two, leading to $\Delta_{\rm loc} \approx 2 D_{\rm min}$ (much as in the elementary theory of level repulsion at degeneracy).   

\subsection{The Expectation Value of the Direct Coupling}
\label{sec:LaplaceMethod}

The last parameter which remains to be estimated, is the expectation value of the direct in-out coupling, $\overline{V}$. Rather than obeying Gaussian statistics such as the coupling between any other two sites of the network, $V$ is governed by so-called {\em extreme value statistics} \cite{deHaan2007extreme}. This is implicitly imposed by construction, since we defined $V = \min_{i}\abs{H_{i, N-i+1}}$, which is the smallest number, in absolute value, of a sample of $N/2$ normally distributed stochastic variables (the Hamiltonian components $H_{i, N-i+1}$). To calculate $\overline{V}$, we start by introducing a method to obtain the distribution of $V$, which we introduce in a general framework and subsequently apply to our specific problem.

To begin with, let $X_1,\dots, X_{n}$ be a sample of $n$ {\em independent, identically distributed} stochastic variables, and denote $m=\min_{k \in \{1,\dots n\}} X_k$. We are now interested in the probability density $P_m(x)=P(m=x).$ To obtain this function, we consider the {\em cumulative distribution function} (CDF) of $m$, $F_m(x)=P(m \leqslant x).$ Since $m$ is the minimum \begin{equation}\label{eq:CDFm}\begin{split}
F_m(x)=P(m \leqslant x)&=1-\prod^n_{k=1}P(X_k>x)\\&=1-\prod^n_{k=1}(1-P(X_k\leqslant x))\\&=1-\left(1-F(x)\right)^n
\end{split}\end{equation}
where $F(x)$ is the CDF of $X_k$. Now the probability density $P_m(x)$ can be obtained as\begin{equation}\label{eq:PDFm}
P_m(x)=\frac{d F_m(x)}{d x}=1-\frac{d}{dx}\left(1-F(x)\right)^n,
\end{equation}
which is seen to strongly depend on the sample size $n$.

In the present case we are dealing with $X_k=\abs{H_{k,N-k+1}}$ and $H_{k,N-k+1} \sim \mathcal{N}\left(0, \frac{2 \xi^2}{N}\right)$ (recall (\ref{eq:ProbCGOEEven})), what implies that $\abs{H_{k,N-k+1}}$ is a half-normal distribution \cite{Leone:1961aa}, therefore the CDF is given by  \begin{equation}\begin{split}
F_{\abs{H_{k,N-k+1}}}\left(x\right)&= \frac{2}{\sqrt{\pi}}\int_{0}^x {\rm d}\abs{H_{k,N-k+1}}~e^{-\left(\frac{\sqrt{N} \abs{H_{i,N-i+1}}}{2\xi} \right)^2}\\&=\text{erf}\left(\frac{\sqrt{N} x}{2\xi}\right),
\end{split}\end{equation}
where ${\rm erf}(x)$ denotes the error function \cite{abr65}.
By using this result and $n=\frac{N}{2}$ in (\ref{eq:CDFm}) and (\ref{eq:PDFm}), we obtain that the probability density of the minimal coupling $V$ is given by \begin{equation}\label{eq:DistSmallV}
P(V)=\frac{e^{-\frac{N V^2}{4 \xi ^2}} N^{3/2} \left(\text{erfc}\left(\frac{\sqrt{N} V}{2 \xi }\right)\right)^{\frac{N}{2}-1}}{2 \sqrt{\pi } \xi },
\end{equation}
with ${\rm erfc(x)}$ the complementary error function, which is given by ${\rm erfc(x)}=1-{\rm erf(x)}$ \cite{abr65}.\\

From these results, $\overline{V}$ is now inferred as
\begin{equation}\begin{split}\label{eq:NiftyAveraging}
\overline{V}&=\int^{\infty}_{0}{\rm d}V \text{ } P(V)V\\&=\int^{\infty}_{0}{\rm d}V \text{ } \frac{e^{-\frac{N V^2}{4 \xi ^2}} N^{3/2} \left(\text{erfc}\left(\frac{\sqrt{N} V}{2 \xi }\right)\right)^{\frac{N}{2}-1}}{2 \sqrt{\pi } \xi } V.
\end{split}
\end{equation}
With the change of variable \begin{equation}V'=\frac{\sqrt{N} V}{2 \xi },\end{equation}  the right hand side of (\ref{eq:NiftyAveraging}) turns into\begin{equation}\label{eq:bla}
\frac{2 \xi \sqrt{N} }{\sqrt{\pi } }\int^{\infty}_{0}{\rm d}V'\text{ } e^{-{V'}^2}  \left(\text{erfc}\left(V'\right)\right)^{\frac{N}{2}-1} V'.
\end{equation} 
Since we are interested in the behavior for large $N$, we have $N/2-1 \approx N/2$. We now apply Laplace's method \cite{Laplace:1986aa}, and thus need to define a function $f$ such that  \begin{equation}\int^{\infty}_{0}{\rm d}V'\text{ } e^{-{V'}^2}  \left(\text{erfc}\left(V'\right)\right)^{\frac{N}{2}-1} V'=\int^{\infty}_{0} \text{d}V' \exp\left(N f(V')\right).\end{equation}
It is straightforward to check that \begin{equation}\label{eq:functionf}
f(V')=-\frac{{V'}^2}{N}+  \left(\frac{1}{2}-\frac{1}{N}\right)\log\left(\text{erfc}\left(V'\right)\right) + \frac{1}{N}\log V' \end{equation} is a suitable choice. In order to apply Laplace's method, we need to find that $V_0$ for which $f$ is extremal, hence $f'(V_0)=0$. By straightforward calculation of the derivative of (\ref{eq:functionf}) we find \begin{equation}\label{eq:functiondf}
f'(V')=\frac{1}{N V'}-\frac{2 V'}{N}-\left(1-\frac{2}{N}\right)\frac{e^{-{V'}^2}}{\sqrt{\pi } \text{erfc}(V')},
\end{equation}
what only allows for an implicit expression for $V_0$. We can however get an explicit result by the following approximation: As the maximum of $f(V')$ is achieved for $V_0\ll 1$, we can expand $e^{-{V'}^2}$ and $\text{erfc(V')}$ around $V' \approx 0,$ in order to obtain a tractable approximation for $f(V')$. This expansion yields \begin{equation}\label{ApproxErf}
\frac{e^{-{V'}^2}}{\sqrt{\pi } \text{erfc}(V')}= \frac{1-{V'}^2 +\frac{1}{2}{V'}^4-\dots}{\sqrt{\pi}(1 -2V' +\frac{2}{3}{V'}^3+\dots)}\approx \frac{1}{\sqrt{\pi}}.
\end{equation}
Even though this is a rough approximation, the corrections due to higher orders are negligible for large $N$ --- numerical evaluation of  (\ref{eq:NiftyAveraging}) shows that, even for $N=10$, the exact results are very well approximated by (\ref{ApproxErf}).

With the low order approximation of (\ref{ApproxErf}), $f'(V_0)=0$ is satisfied for \begin{equation}V_0 \approx \frac{\sqrt{N^2+8 \pi }-N}{4 \sqrt{\pi }}\approx \frac{\sqrt{\pi}}{N}\left(1+\frac{2}{N}\right),\end{equation} and Laplace's method now tells us that {\begin{equation}
\int^{\infty}_{0} \text{d}V' \exp\left(N f(V')\right)\approx  \text{e}^{Nf(V_0)} \sqrt{\frac{2\pi}{N \abs{f''(V_0)}}},
\end{equation}
leading to the final result 
 \begin{equation*}\tag{\ref{eq:vbarFinal}}\overline{V}\approx\frac{2 \pi \xi}{e N \sqrt{N/2-1}},\end{equation*}
which we already anticipated in Section \ref{sec:ObtDist} above, to obtain the transfer time distribution (\ref{eq:Dist}).

\section{Simulations for Random Hamiltonians}\label{sec:sim}

Having completed the derivation of the analytical predictions of our constrained (by centrosymmetry and dominant doublet assumption) RMT model for efficient transport on random graphs, we now test these predictions against numerical simulations. We sample random Hamiltonians  from the GOE, with centrosymmetry imposed as an extra constraint. After diagonalisation of each of these Hamiltonians, we post-select those which exhibit a dominant doublet with weight $\alpha,$ as defined in (\ref{eq:domdubalpha}).
Then, from the thus constructed RMT-ensemble, we numerically derive $\mathcal{P}_H$, with $t$ the earliest point in time for which $\abs{\braket{\text{out}}{\phi(t)}}^2=\mathcal{P}_H$.\\

%
%

\begin{figure*}

\includegraphics[width=0.49\textwidth]{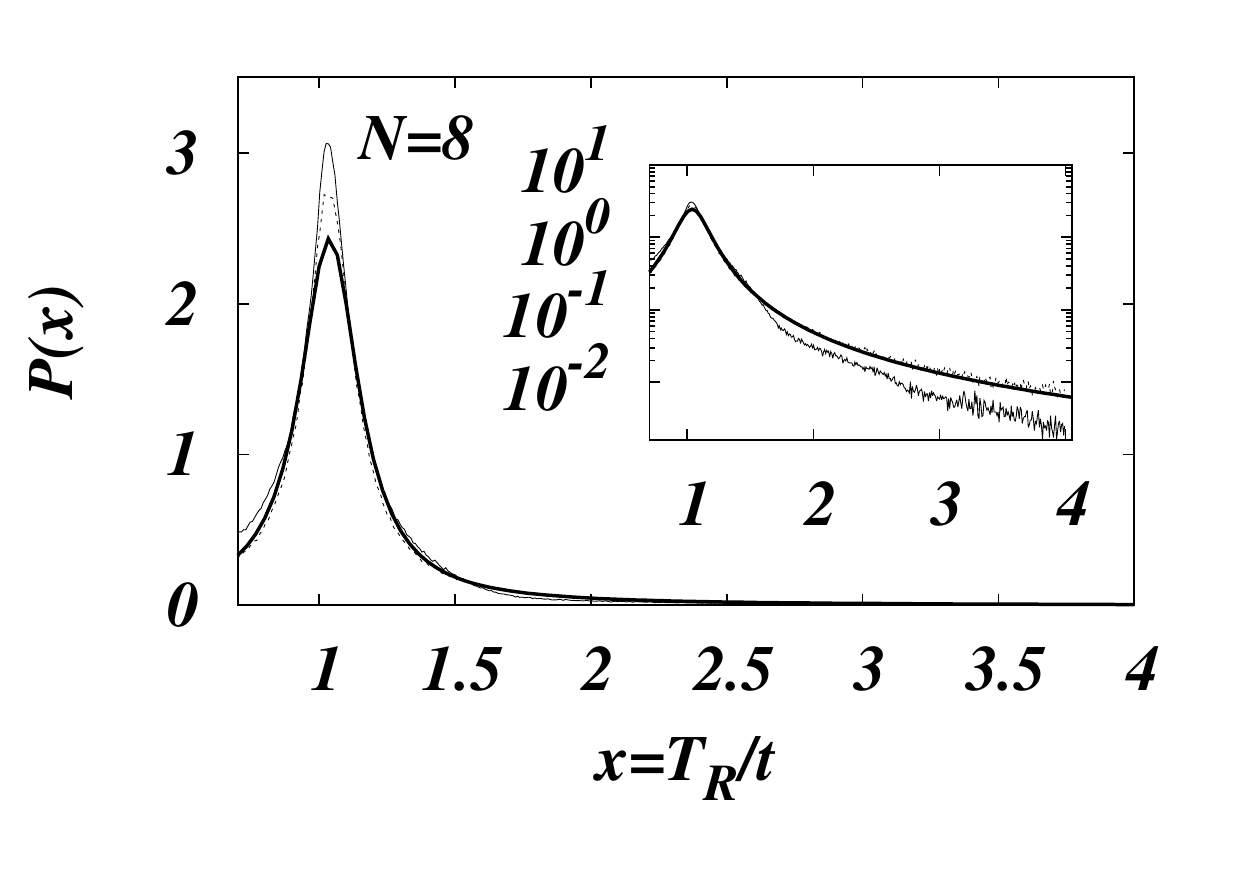}
\includegraphics[width=0.49\textwidth]{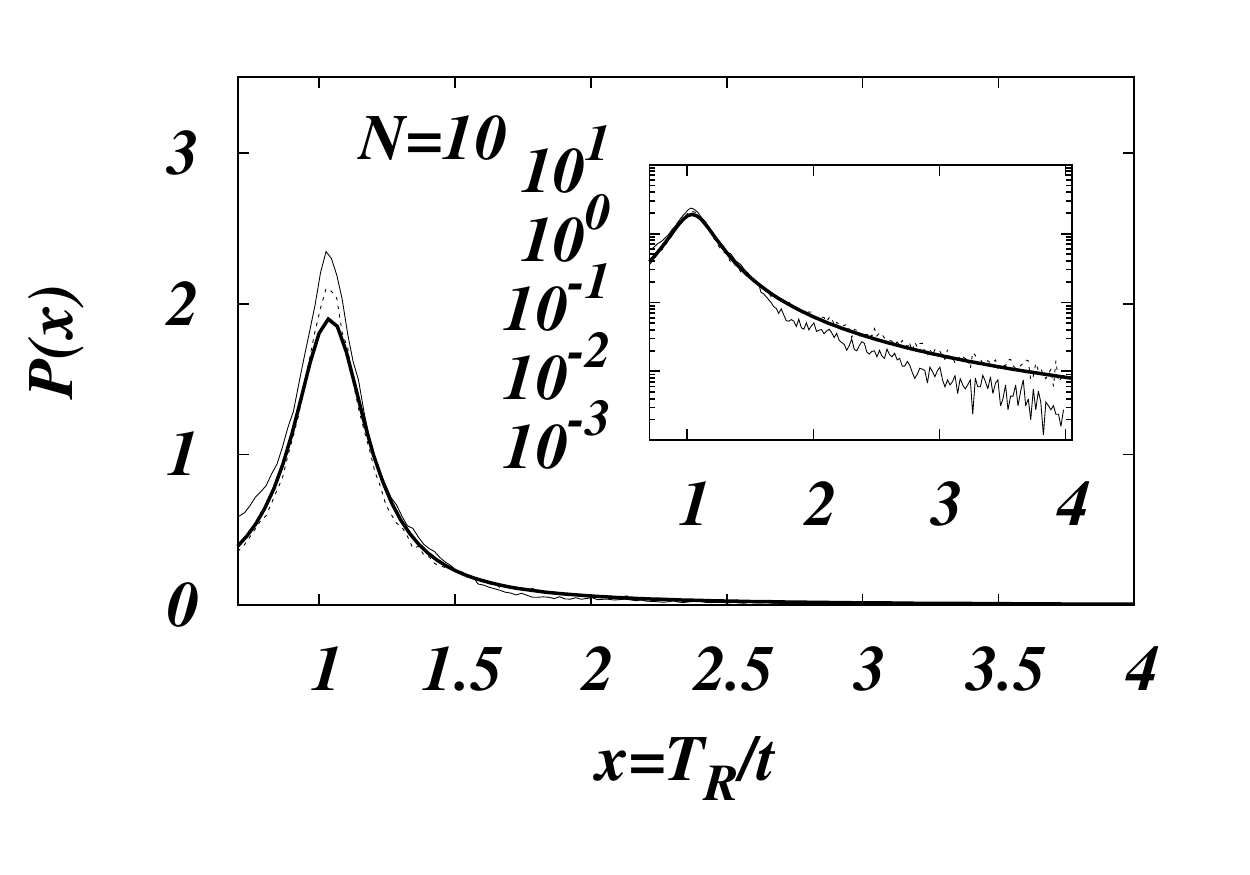}
\includegraphics[width=0.49\textwidth]{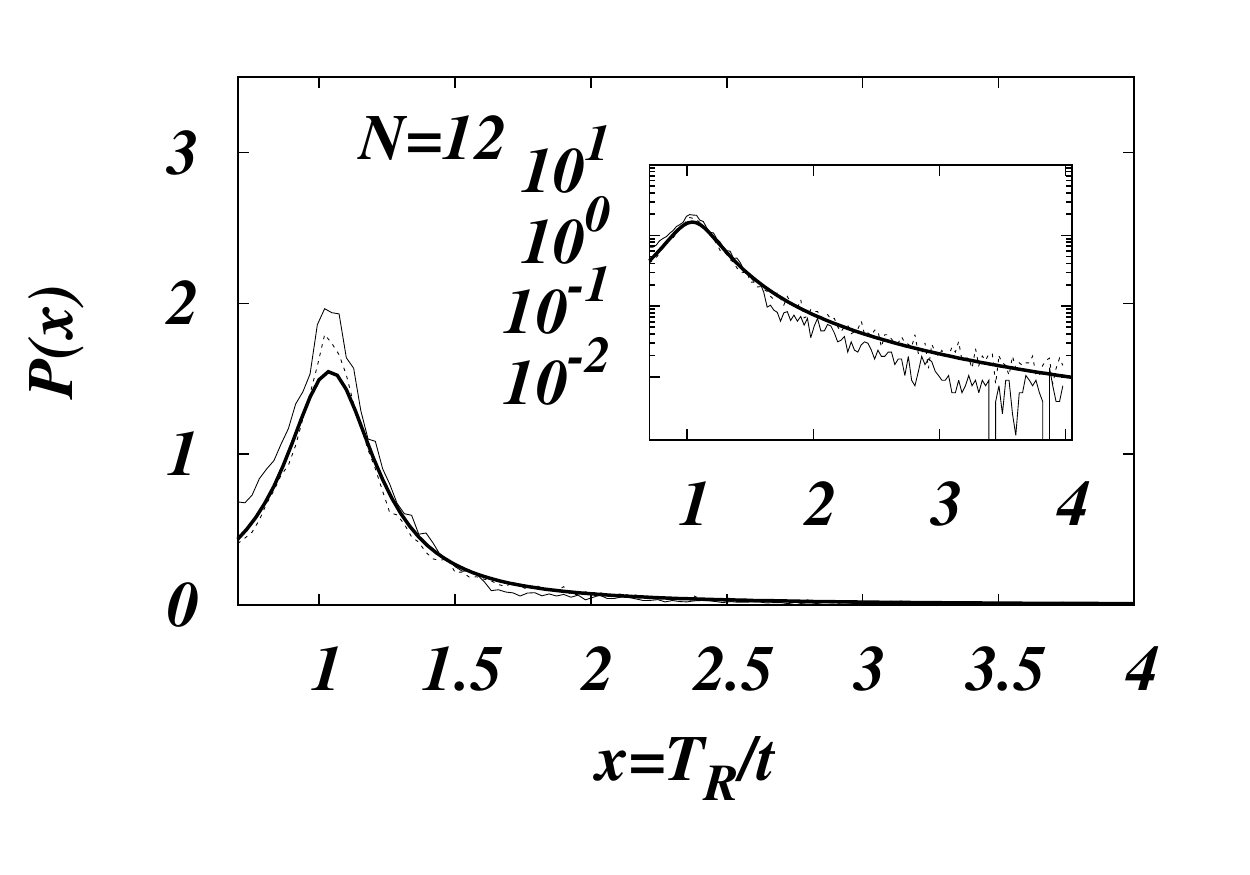}
\includegraphics[width=0.49\textwidth]{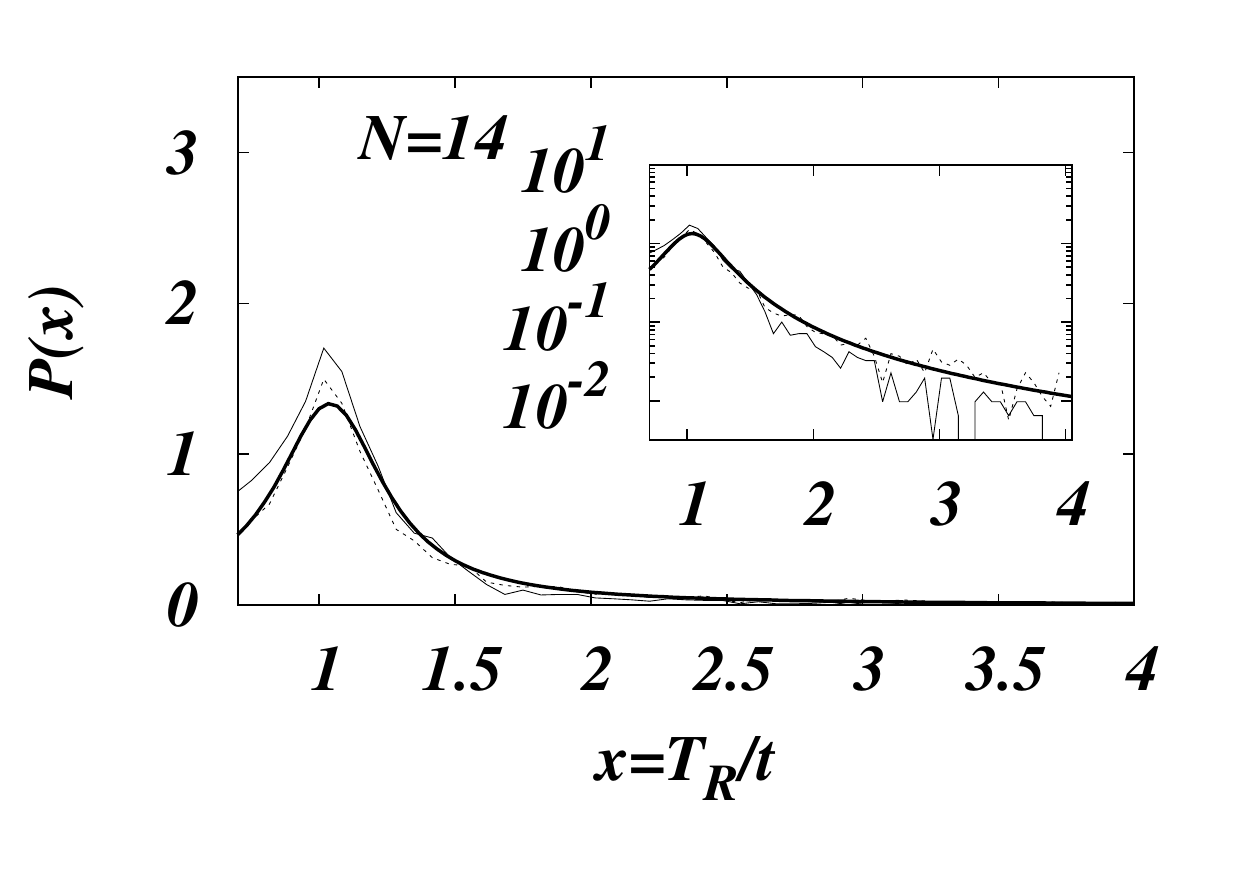}

\caption{Histograms of the simulated inverse transfer time $T_R/t$ (thin solid lines) across fully connected random networks of variable size $N$, 
and of $\abs{E^+-E^-}/2 V$ (dashed lines), 
together with the theoretical distribution (\ref{eq:Dist}) (thick solid line). The parameters $\xi=2$ and $\alpha=0.95$ are fixed for every realisation. The value $\overline{\norm{\mathcal{V}}^2} \approx 0.31$ is extracted from the simulations for each value of $N$. The simulations only consider a time window $[0,1.7 T_R]$, therefore the minimum value of the inverse transfer time is given by 
$T_R/t=(1.7)^{-1}$. The inset stresses the agreement between the theoretically predicted algebraic tail (thick solid line) and the $\abs{E^+-E^-}/2 V$ histogram (dashed line). The histogram for $T_R/t$ (thin solid line) slightly deviates from the other two curves because the quasi-periodicity of the dynamics suppresses the tail of the distribution (see text).}
\label{fig:time}
\end{figure*}

To start with, Figure \ref{fig:time} shows the transfer time distribution for different network sizes $N$ --- at fixed spatial density (remember our discussion of (\ref{eq:Asympt1}, \ref{eq:Asympt2}) above), with a comparison between numerical data (thin solid line) and the analytical prediction (\ref{eq:Dist}) (thick solid line). There are {\em no} fitting parameters; the average coupling strength $\overline{\norm{\mathcal{V}}^2}$ is directly extracted from the statistical sample, whereas the dominant doublet strength $\alpha=0.95$ and spectral density $\xi=2$ (in units of mean level spacing) are fixed a priori for all realisations. 

The overall comparison of numerical data and analytical prediction is very satisfactory. In particular, the distribution also exhibits the trend predicted by (\ref{eq:Dist}, \ref{eq:Asympt1}) for increasing $N$: As $N$ grows, the height of the maximum of the distribution at $T_R/t \approx 1$, controlled by $s_0$ (see (\ref{eq:Dist})) decreases, and the algebraic tail with $T_R/t \gg 1$ grows fatter, as anticipated by (\ref{eq:Asympt1}). Indeed, the numerical data confirm the predicted scaling of $s_0$ and $P(T_R/t > 1)$ with $N^{-1}$, as spelled out by Fig.~\ref{fig:Scaling}.

However, closer scrutiny of the displayed distributions for larger values of $T_R/t$ (see the insets of Fig.~\ref{fig:time}) suggests an apparent discrepancy between numerics and analytical prediction: The numerical data appear to drop faster with increasing $T_R/t$ than expected from (\ref{eq:Dist}), which was derived from the statistics of the first passage time (\ref{eq:transfertime}). It turns out that this is an effect caused by the quasi-periodic oscillation between the input and the 
output site. If, \eg~$T_R/t > 3$, the excitation will localise on the output site three times during the benchmark time interval $[0,T_R)$. Since, however, the dynamics is in general quasi-periodic, rather than periodic (note that this is a consequence of the transient population of the bulk sites, which is neglected in the approximate expression (\ref{eq:transfertime}) for the transfer time in terms of the dominant doublet splitting), the largest value of $\abs{\braket{\text{out}}{\phi(t)}}^2$ within the considered time window may only be achieved after multiple periods. Even though the theoretical value of $t$ is relatively small, the simulation may pick up a later point in time, thus giving  a smaller weight to large values of $T_R/t$ in the histograms of Fig.~\ref{fig:time}. One incident of this scenario is shown in Figure \ref{fig:FastDynamics}.

Indeed, direct comparison of the time scale (\ref{eq:transfertime}) given by the numerically sampled doublet splitting (\ref{eq:blub}) (rather than of $t$ as inferred from direct propagation of the associated unitary generated by $H$) leads to perfect agreement  in particular of the asymptotic behaviour of the distribution with the analytical prediction, as evident from comparison of the dotted and full curves in Fig.~\ref{fig:time}. The dominant doublet mechanism is thus impressively confirmed, with an asymptotic behaviour inherited from the statistics of the level shifts $\Delta s$, induced by the interaction with the network's bulk sites.

\begin{figure}[t] 
   \centering
  	\includegraphics[width=0.49\textwidth]{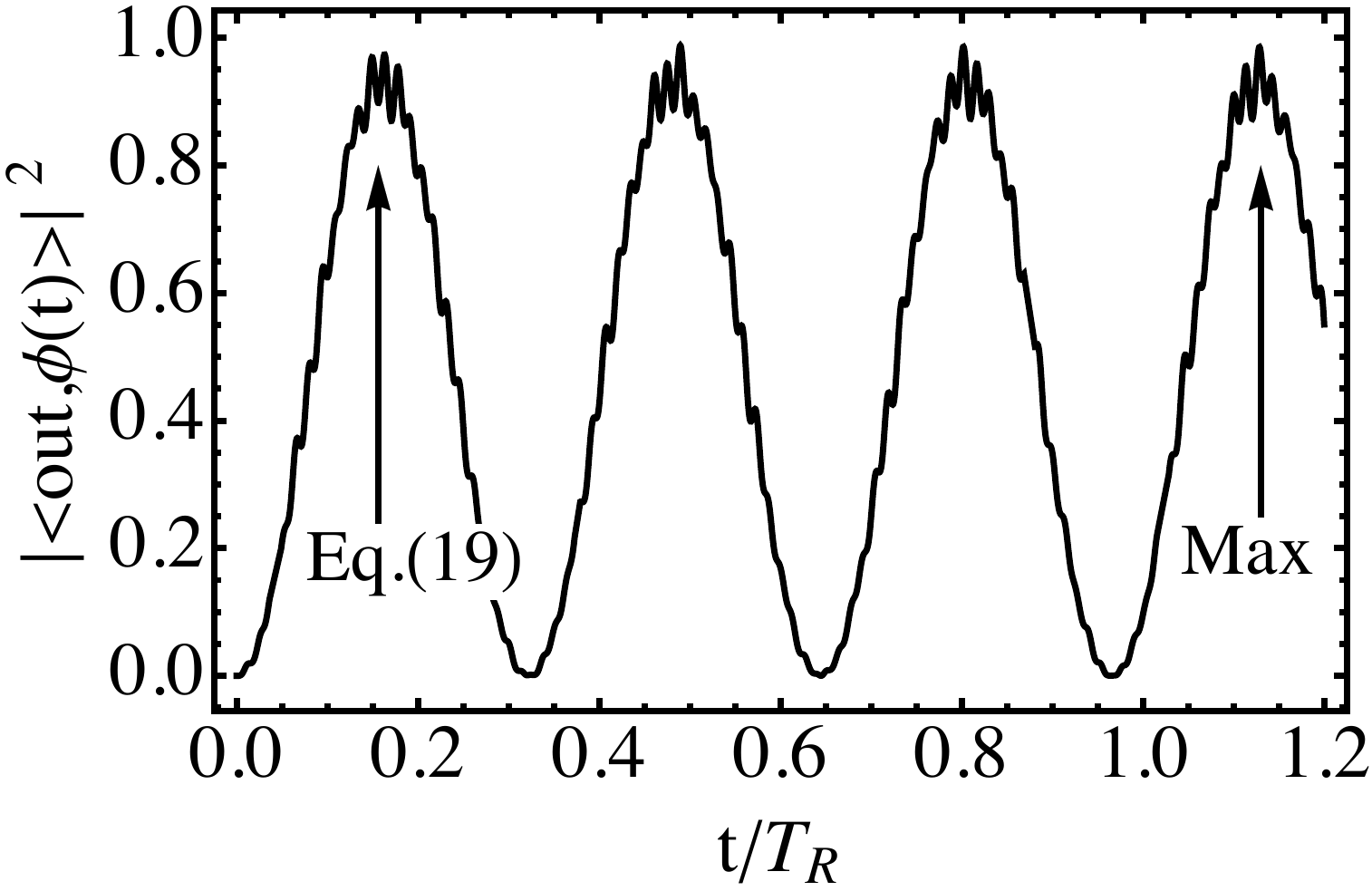}
  \caption{Probability $\abs{\braket{\text{out}}{\phi(t)}}^2$ to find the excitation at the output site, for a single realisation of the Hamiltonian (\ref{eq:Matrix}). There are multiple strong localisations at the output site within $[0,T_R)$. High frequency oscillations show that the dynamics is quasi periodic rather than periodic.}
   \label{fig:FastDynamics}
\end{figure}

Having achieved an excellent understanding of the transfer time distribution of centrosymmetric random graphs with dominant doublet, we still need to verify that they indeed also generate large transfer probabilities ${\cal P}_H\geqslant 2\alpha -1 \approx 0.9$, for the here chosen  dominant doublet strength $\alpha=0.95$. This is done in Fig. \ref{fig:eff}, for the same model parameters as in Fig. \ref{fig:time}, and in comparison to the efficiency distribution of unconstrained or just centrosymmetric random graphs (obeying GOE statistics). For different values of $N$, the figure provides an impressive illustration of the here suggested {\em design principles}: centrosymmetry alone already induces a very tangible shift of the average value of the transfer efficiency to much enhanced values, though fails to concentrate the distribution to values close to one. This is unambiguously achieved by the dominant doublet constraint (a generalisation of the CAT mechanism), and in perfect agreement with our predictions.

\begin{figure*}[t]
\centering
\includegraphics[width=0.49\textwidth]{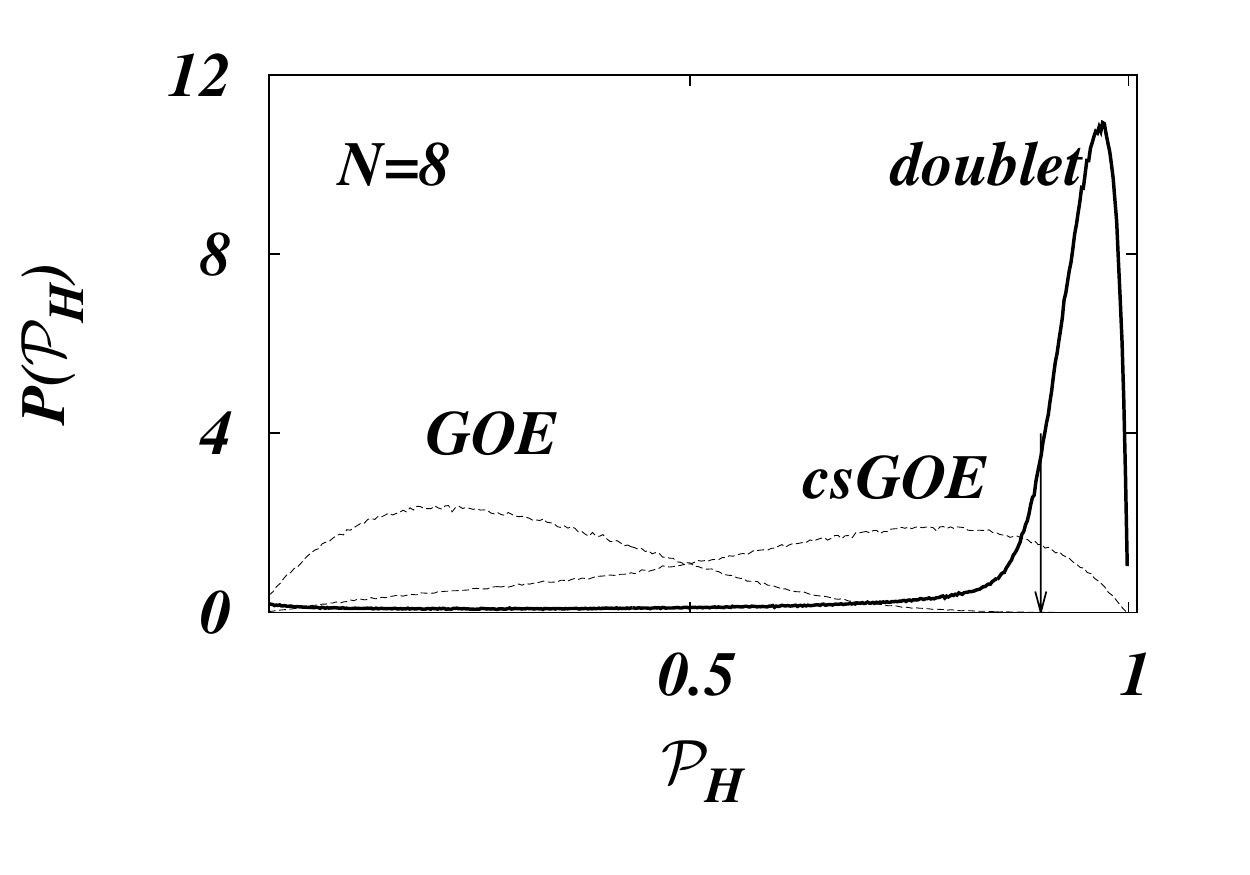}
\includegraphics[width=0.49\textwidth]{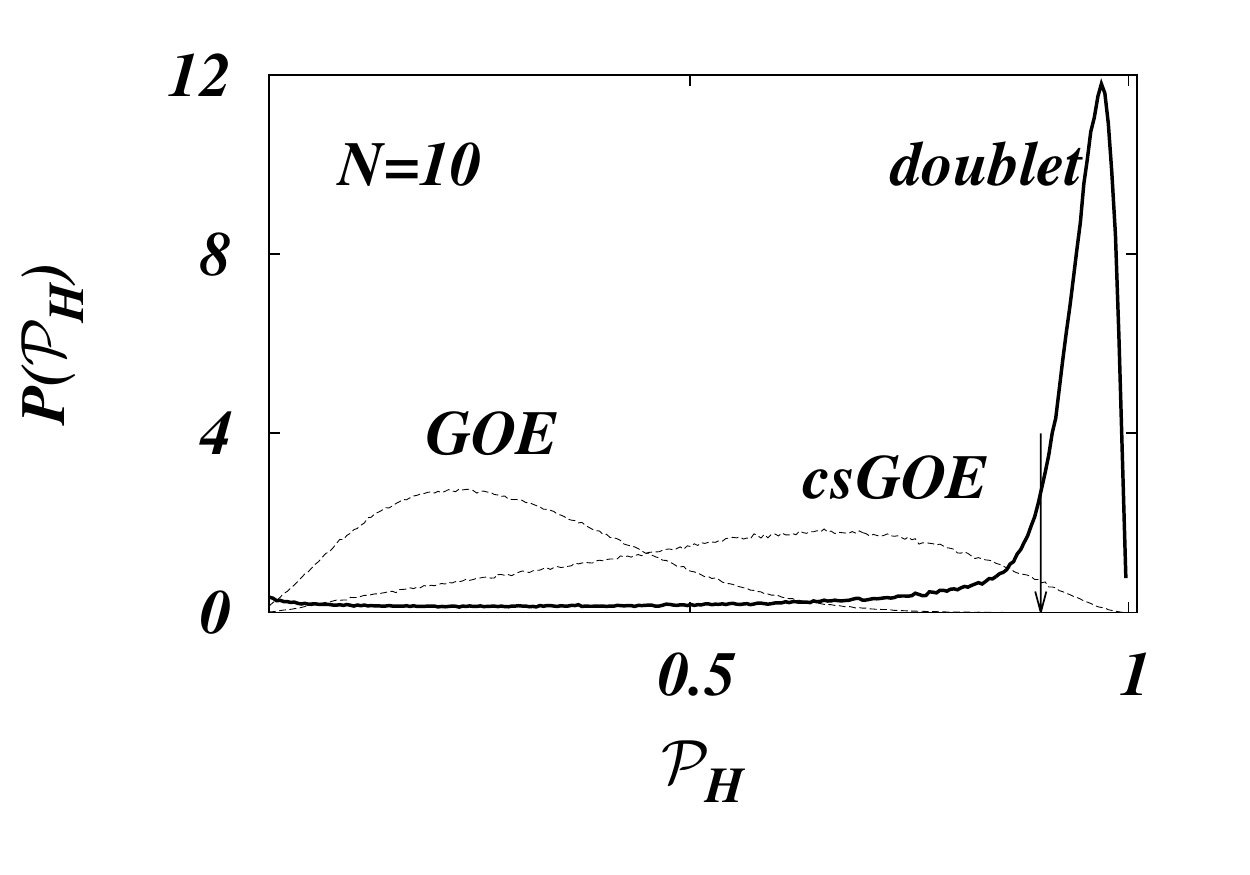}
\includegraphics[width=0.49\textwidth]{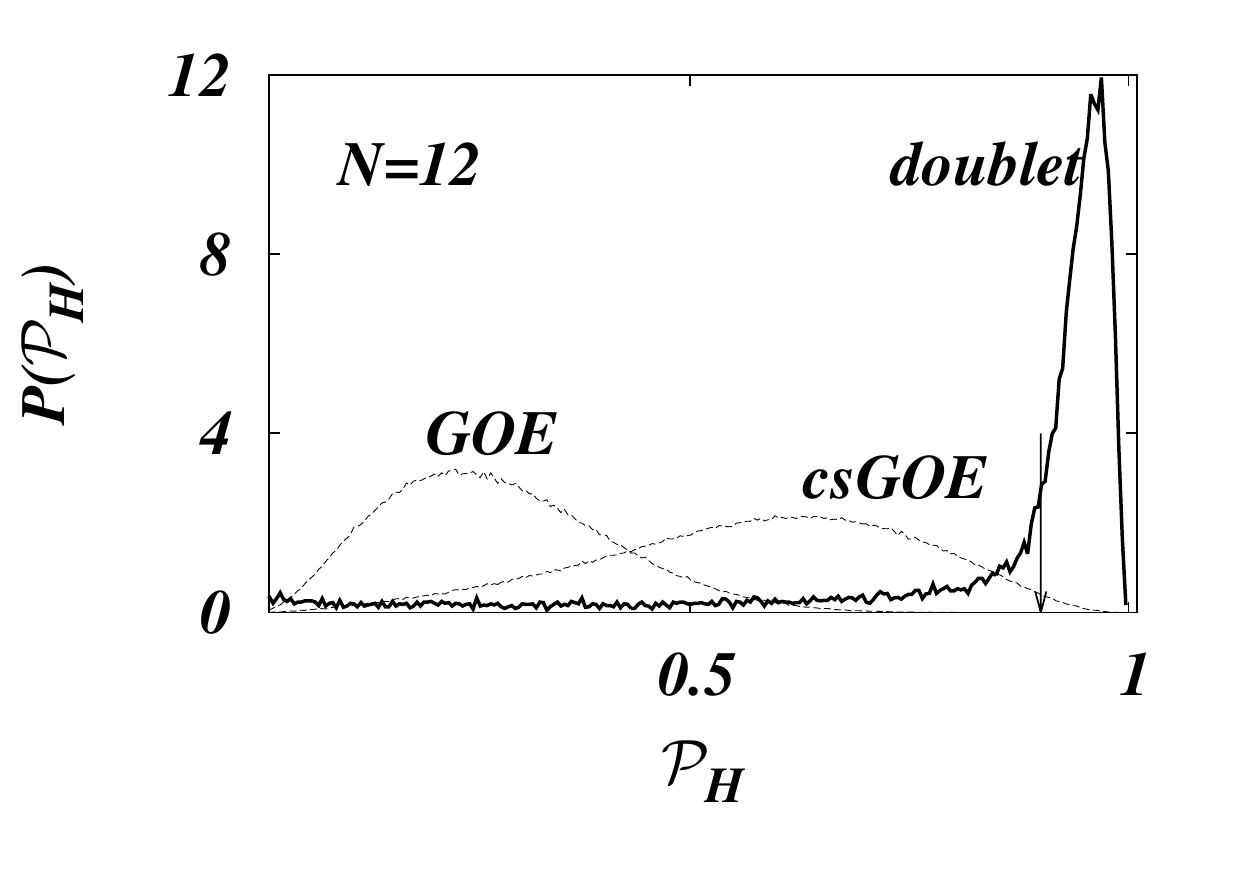}
\includegraphics[width=0.49\textwidth]{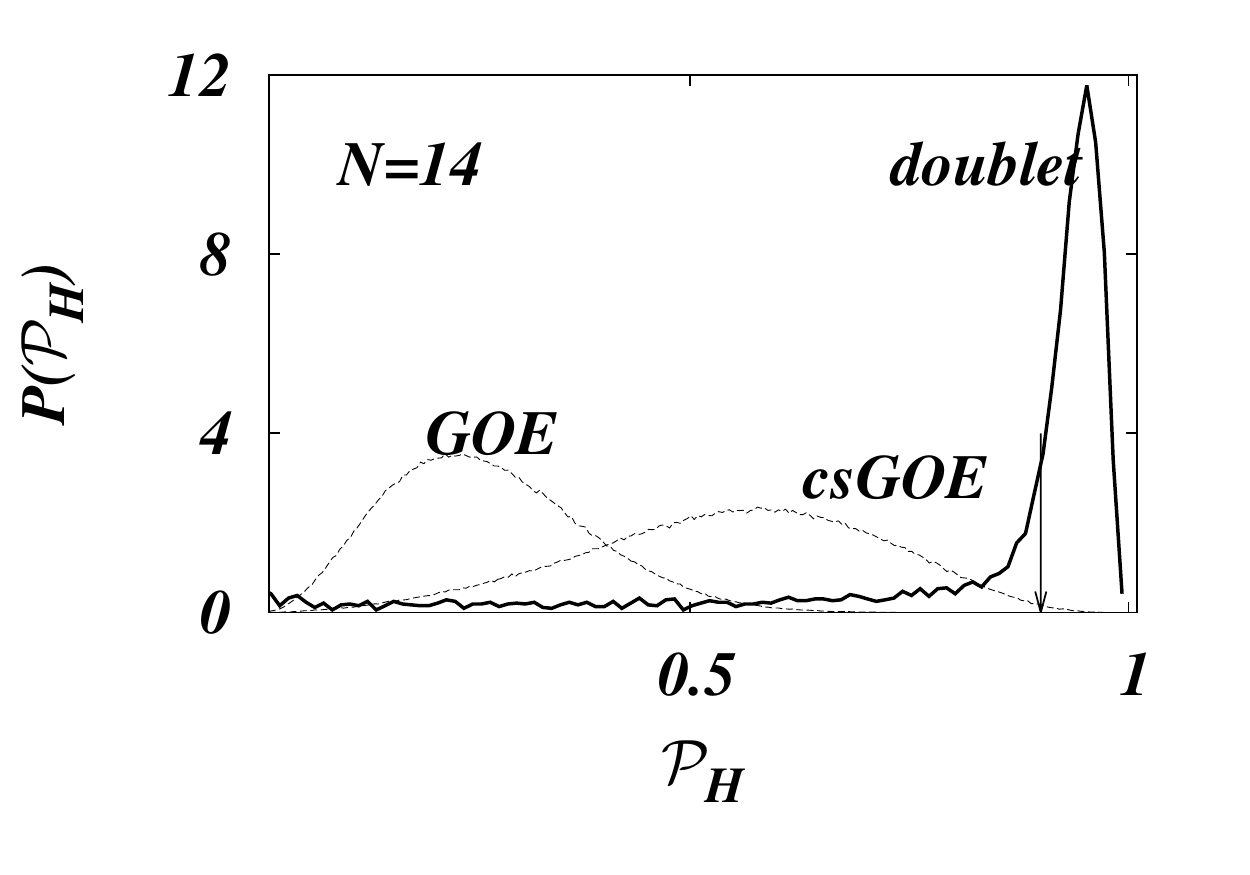}
\caption{Distribution of the transfer efficiency $\mathcal{P}_H$, eq.~(\ref{eq:eff}), for variable network size $N$ and three Hamiltonian ensembles: GOE, GOE with centrosymmetry, and GOE with centrosymmetry and dominant doublet. $\mathcal{P}_H = 2\alpha-1$ is indicated by the arrow. The control parameters in (\ref{eq:ProbCGOEEven},\ref{eq:domdubalpha}) are set to $\xi=2$ and $\alpha=0.95$.}\label{fig:eff}
\end{figure*}

As for those incidents in Figure \ref{fig:eff} where $ \mathcal{P}_H < 2\alpha - 1$, despite the presence of a dominant doublet, these typically are due to network conformations where $T_R < t$, \ie, where the transport is too slow to be efficient. On the other hand, there are also some realisations (such as shown in Figure \ref{fig:Dynamics}) where $ \mathcal{P}_H > 2\alpha - 1$ even though $T_R < t$. As a matter of fact, there is no obvious one-to-one relation between the first passage time distribution and the efficiency distribution what, however, leaves our overall picture of the transport mechanism fully intact. Also, as $N$ is increased, it might seem that more structure emerges in the region $ \mathcal{P}_H < 2\alpha - 1$. This, however, is just statistical noise: Because of the post-selection, together with the strongly decreasing density of dominant doublets in the ensemble (Appendix \ref{sec:FindingDoublets}), it is difficult to acquire a lot of statistics for $N=14$.

\begin{figure}
\centering
\includegraphics[width=0.49\textwidth]{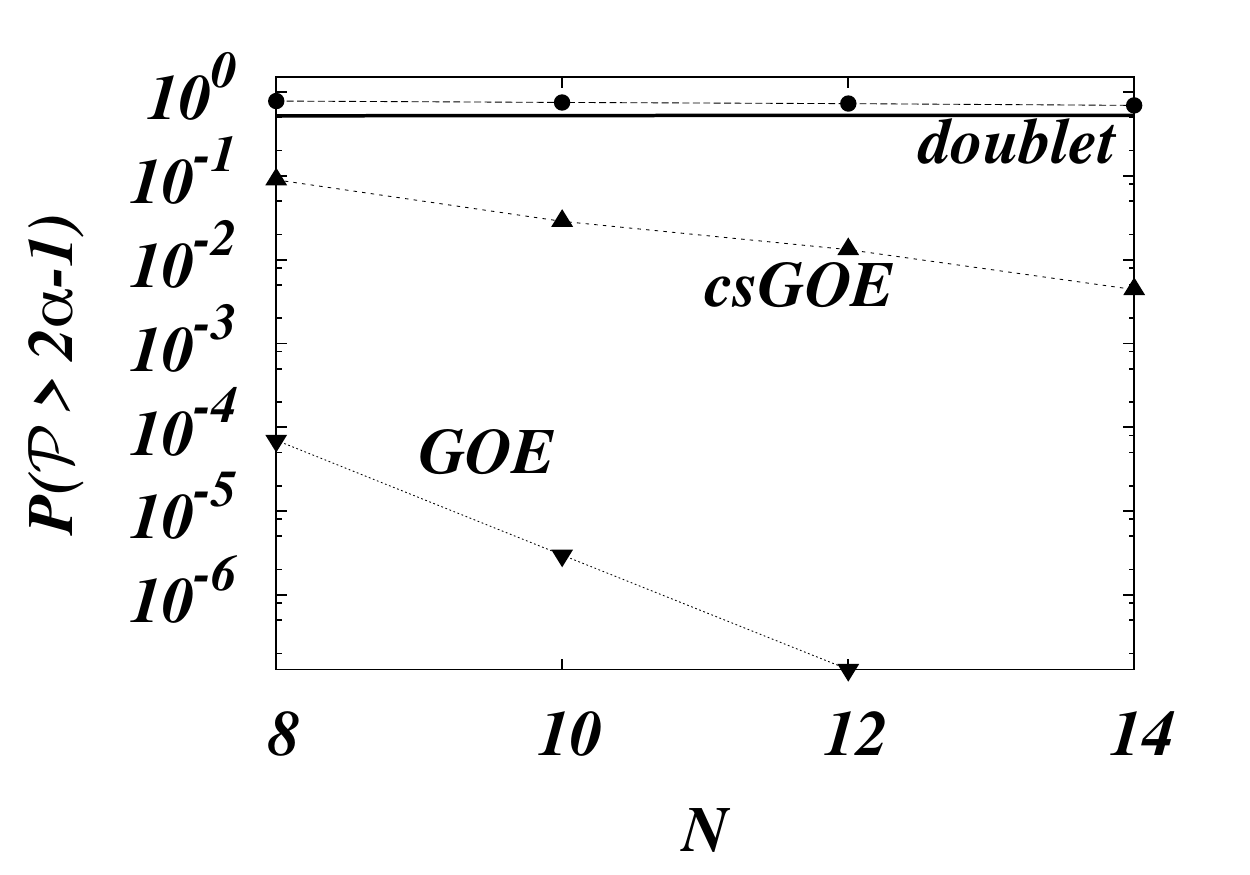}
\caption{Density $P(\mathcal{P}_H> 2\alpha-1)$ of efficient network realisations,  as a function of the network size $N$,
for three different ensembles (GOE, GOE with centrosymmetry, GOE with centrosymmetry and dominant doublet).  
The theoretical curve for $P(t < T_R)$ (solid line), eq.~(\ref{eq:ProbLargerAlpha2}), forms a lower bound to the dominant doublet ensemble, as expected (see text). The GOE curve is cut off at $N=12$, since it takes too long to sample a sufficient amount of data for larger values of $N$. The control parameters in (\ref{eq:ProbCGOEEven},\ref{eq:domdubalpha}) are $\alpha=0.95$ and $\xi=2$.}
\label{fig:Scaling}
\end{figure}

Let us finally extract from Fig. \ref{fig:eff} the probability to achieve transfer efficiencies ${\cal P}_H>2\alpha-1$, what is simply done by integrating over the corresponding interval of the histograms, for the different ensembles considered. The result displayed in Fig. \ref{fig:Scaling} is yet  another impressive demonstration of the effectiveness of centrosymmetry and dominant doublet as robust design principles.
Also note that the result for the dominant doublet ensemble confirms  the estimate (\ref{eq:ProbLargerAlpha2}): Since $T_R/t>1$ guarantees ${\cal P}_H>2\alpha-1$, while the inverse is not true (remember Fig. \ref{fig:Dynamics}), (\ref{eq:ProbLargerAlpha2}) defines a {\em lower bound} for $P({\cal P}_H>2\alpha-1),$ as nicely spelled out by the comparison in Fig. \ref{fig:Scaling}.

\section{Discussion and Conclusions}

We described a general mechanism that gives rise to fast and efficient quantum transport on finite, 
disordered networks. The mechanism rests on two crucial ingredients: 
The first is the {\em centrosymmetry} of the underlying Hamiltonian, which renders the Hamiltonian block-diagonal in the eigenbasis of the exchange matrix --- the symmetry operator. The second ingredient is a {\em dominant doublet}, that 
ensures a firm control of the transport properties' statistics, under the coupling to random (or chaotic --- recall the original motivation of the CAT mechanism \cite{PhysRevE.50.145}) states which 
assist the transport. The statistics of the transfer efficiencies and times as 
shown in Figs. \ref{fig:time} and \ref{fig:eff} only depend on the intermediate network sites' density of states $\xi$, and on the average coupling
strength $\overline{\norm{\mathcal{V}}^2}$ of the in- and output-sites to the network. These are macroscopically controllable parameters.
On the one hand, this means that coherence effects survive simply by stabilising these {\em properties of the ensemble}.
On the other hand, if such stabilisation is possible, one could also imagine controlling transport properties according to the specific needs, simply by controlling these ensemble properties such as the density of states and the typical coupling to the intermediate sites.\\

The key point of our contribution is to treat near-optimal transport in a context of disorder physics, where we do not strive to avoid disorder altogether, but rather incorporate it in a constructive way. By no means do we wish to control as many degrees of freedom as possible, as hardwired small-scale 
structures are unavoidably perturbed by omnipresent fluctuations. Rather we provide a framework that optimally controls few coarse grained quantities --- only constrained by 
the above design principles  ---, whereas microscopic details may remain subject to disorder/fluctuations. Our handle of control tunes the statistical properties of the transfer efficiency in the sense that it controls the shape of the distributions in Figs. \ref{fig:time} and \ref{fig:eff}. Moreover, the transfer time distribution, Fig. \ref{fig:time}, is a Cauchy distribution, which, as its possibly most important feature, has an algebraic (fat) tail, guaranteeing that transfer times which are shorter than the Rabi time occur in a relatively large fraction of network realisations. In particular, there is a non-negligible probability for dramatic speed-up (by more than an order of magnitude)  
of the excitation transfer. 

Recently, other works concerning ensemble approaches to efficient transport in complex quantum systems have been presented \cite{Mostarda:2013ab,Mostarda:2013aa}, where, by randomly sampling networks of dipoles, realisations leading to efficient transfer are identified. When analysing these networks, one mainly encounters centrosymmetric structures \cite{Mostarda:2013ab}. As these efficient realisations are further investigated, \cite{Mostarda:2013ab,Mostarda:2013aa} find that typically only a subset of the network sites are significantly populated during the transport, which is a consequence of Hamiltonian eigenvector localisation on these network sites. Although \cite{Mostarda:2013ab,Mostarda:2013aa} encounter different possible {\em backbone structures}, typically containing four sites or more, these results are strongly reminiscent of our dominant doublet. In other words, one might say that the dominant doublet is a specific --- and (see above) analytically tractable --- type of backbone structure. We expect that the more complex backbones of \cite{Mostarda:2013ab,Mostarda:2013aa} can be incorporated into a framework similar to the one which is presented in our present contribution, by adopting models comparable to what is known as {\em Resonance Assisted Tunnelling} in the quantum chaos literature \cite{PhysRevLett.87.064101,Brodier:2002aa}.\\

Finally, even though our work is originally inspired by recent developments in photobiology, as we stressed in detail in \cite{Walschaers:2013aa,Zech:2014aa}, one might think of various other fields where ensemble approaches to quantum transport are rapidly gaining relevance. More specifically, the realm of quantum computation harbours several ideas that relate computational problems to quantum walks \cite{hein2009,PhysRevLett.102.180501,PhysRevA.58.915}, thus relating quantum computation to complex networks. On the other hand, random matrix models have been successfully applied in the study of adiabatic quantum computation \cite{PhysRevA.71.032330}. More recently, in the broad discussion on quantum effects in {\em D-Wave Two} \cite{bunyk_architectural_2014}, it became clear that random fluctuations and disorder effects must be incorporated in the study of quantum effects in such real systems \cite{PhysRevX.4.021041}. We trust 
that a model as ours, in all its generality, may also enrich this field.\\

{\bf Acknowledgements:} R.M. acknowledges support by the Alexander von
Humboldt Stiftung. M.W. and A.B. are grateful for funding within the DFG Research Unit 760 and for support
through the EU COST Action MP1006. M.W. acknowledges partial funding by the Belgian Inter-university
Attraction Poles Programme P6/02 and FWO Vlaanderen
Project No. G040710N, and by the German National Academic Foundation. 

\appendix

\section{Block Diagonalization of $H$}\label{sec:Block_diagonalization}

The procedure to obtain the block diagonal form is the same as in \cite{Cantoni:1976aa}. We consider $H \in \mathbb{C}^{N\times N}$ to be a symmetric {\em and} centrosymmetric matrix. We now represent $H$ in terms of four matrices $A,B,C,D \in \mathbb{C}^{N/2\times N/2}$, such that \begin{equation}
H=\begin{pmatrix}
A&B\\C&D
\end{pmatrix}.
\end{equation}
This can of course be done for any matrix. Symmetry and centrosymmetry now imply that \begin{equation}B=J'CJ' \qquad \text{and} \qquad D=J'AJ' .\end{equation}
Here $J'$ is an $N/2 \times N/2$ matrix such that the exchange operator $J$ is given by \begin{equation}
J=\begin{pmatrix}0&J'\\J'&0
\end{pmatrix}.
\end{equation}Next, a transformation $K$ is defined such that $K \in \mathbb{C}^{N\times N}$ is orthogonal and given by \begin{equation}\label{eq:K}
K=\frac{1}{\sqrt{2}}\begin{pmatrix}\mathbb{1}_{\frac{N}{2}}&-J'\\\mathbb{1}_{\frac{N}{2}}&J' \end{pmatrix}.
\end{equation}
Therefore it follows from a simple calculation that \begin{equation}\label{eq:StructCentroBlock}
KHK^{T}=\begin{pmatrix}A-J'C & 0\\ 0 & A+J'C\end{pmatrix}.
\end{equation}\\
From this result it is now easy to identify $H^{\pm}= A\pm J'C$ in terms of random matrix quantities: Consider that, since $A_{ij}, (J'C)_{ij} \sim \mathcal{N}\left(0, \frac{\xi^2}{N}\right)$, if $i\neq j$, it follows from basic probability theory that \begin{equation}
(A+J'C)_{ij}=A_{ij}+ {J'C}_{ij} \sim \mathcal{N}\left(0, \frac{2\xi^2}{N}\right), \quad i\neq j.
\end{equation}
Likewise we have
\begin{equation}
(A+J'C)_{ii}=A_{ii}+ {J'C}_{ii} \sim \mathcal{N}\left(0, \frac{4\xi^2}{N}\right).
\end{equation} Since the sum of two symmetric matrices is again symmetric, and since the components are sampled from a Gaussian distribution, these matrices belong to the GOE. In the case of $A-J'C$, there is an extra subtlety because of the minus sign. Here we explicitly use that a Gaussian distribution is symmetric, such that, if $(J'C)_{ij} \sim  \mathcal{N}\left(0, \xi^2/N\right)$, also $-(J'C)_{ij} \sim  \mathcal{N}\left(0, \xi^2/N\right),$ what implies  
\begin{equation}
(A-J'C)_{ij} \sim\begin{cases} \mathcal{N}\left(0, \frac{2\xi^2}{N}\right), \quad &\text{ if } i\neq j.\\  \mathcal{N}\left(0, \frac{4\xi^2}{N}\right), &\text{ if } i=j.\end{cases}
\end{equation}
Consequently, also $A-J'C$ is an $N/2 \times N/2$ GOE matrix.

\section{Finding Dominant Doublets}\label{sec:FindingDoublets}

In Section \ref{sec:model}, we introduced the dominant doublet as a constraint. From a theoretical point of view, it is also interesting to have an idea of the probability that this constraint holds for any GOE matrix with the centrosymmetry constraint.   

At first we quote an interesting result from \cite{Haake:2010aa}, which concerns eigenvectors. We consider an $N \times N$ matrix in the GOE. Without loss of generality, the eigenvectors of a GOE matrix can be taken to be real. Since an eigenvector $\ket{\eta}$ can be mapped onto any other real vector by an orthogonal transformation, every eigenvector occurs with the same probability. The only property that needs to be fixed is the norm. This implies that \begin{equation}
P_{GOE}(\ket{\eta})=C\text{ } \delta\left(1-\sum_{i=1}^ N \eta_i^2\right).
\end{equation}
Here $C$ is a normalization factor. After determining $C$ and integrating out $N-1$ components, we obtain the distribution for $y=\eta_j^2$, where $\eta_j$ is just some component of the eigenvector $\ket{\eta}$. The result is given by
\begin{equation}
\begin{split}\label{eq:EvecCompDist}
P_{GOE}(y)&=\int_{\mathbb{R}^{N}}\prod^ N_{j=1} {\rm d}\eta_i\text{ } \delta\left(y-\eta_1^2\right) P_{GOE}\left(\ket{\eta}\right)\\
&=\frac{1}{\sqrt{\pi}}\frac{\Gamma\left(\frac{N}{2}\right)}{\Gamma\left(\frac{N-1}{2}\right)}\frac{(1-y)^{\frac{N-3}{2}}}{\sqrt{y}}.\\
&= \frac{1}{B\left(\frac{1}{2},\frac{N}{2}-\frac{1}{2} \right)}y^{1/2-1}(1-y)^{N/2-1/2-1}.
\end{split}
\end{equation}
The last step rewrites this function such that $B(a,b)$ denotes a \textit{Beta function} \cite{abr65}. This implies that the $y$ follow a Beta distribution, \begin{equation}\label{eq:beta}
y \sim \text{Beta}\left(\frac{1}{2}, \frac{N}{2}-\frac{1}{2}\right)\ .
\end{equation}
The quantity of interest is the probability that $\ket{+}$ and $\ket{-}$ from (\ref{eq:pm}) form a dominant doublet. In mathematical terms, this is the probability that for both, $H^{+}$ and $H^{-}$ from (\ref{eq:Hblock}), there exists and eigenvector --- denoted $\ket{\tilde{+}}$ and $\ket{\tilde{-}}$, respectively --- such that \begin{equation}\abs{\braket{\tilde{\pm}}{\pm}}^2>\alpha.\end{equation} This quantity is equivalent to defining \begin{equation}\label{eq:someA}
y=\min\left( \max_{i}\abs{\braket{\eta_i}{+}}^2, \max_{i}\abs{\braket{\eta_i}{-}}^2\right)\ .
\end{equation}
where $\{\ket{\eta_i}\}$ denotes the set of eigenvectors of $H$. Eventually, $\braket{\eta_i}{+}$ and $\braket{\eta_i}{-}$ in (\ref{eq:someA}) are just components of the eigenvector in the eigenbasis of $J$. Remembering (\ref{eq:beta}), we know that for GOE matrices, these components are distributed according to a Beta distribution. As currently we consider Hamiltonians of the form (\ref{eq:Hblock}), we have to treat $H^+$ and $H^-$ as two independent GOE matrices. This implies that the probability that a component $y_i$ is smaller than $\alpha$ is given by \begin{equation}
P\left(\abs{\braket{\eta_i}{\pm}}^2 \leqslant \alpha\right) = I_{\alpha}\left(\frac{1}{2}, \frac{N}{4}-\frac{1}{2} \right),
\end{equation}
where $I_{\alpha}$ denotes the regularized Beta function \cite{abr65}. Since our interest lies in the maximum of $\abs{\braket{\eta_i}{+}}^2$, we can simply follow an approach similar to the one presented in Section \ref{sec:LaplaceMethod}, to obtain \begin{equation}\begin{split}
P\left(\max_i \abs{\braket{\eta_i}{\pm}}^2  > \alpha\right ) &= 1- \prod_i P\left( \abs{\braket{\eta_i}{\pm}}^2_i \leqslant \alpha\right)\\& = 1-\left( I_{\alpha}\left(\frac{1}{2}, \frac{N}{4}-\frac{1}{2} \right)\right)^{N/2}.
\end{split}
\end{equation}
Now that we know the probability for both $\max_i \abs{\braket{\eta_i}{+}}^2>\alpha$ and $\max_i \abs{\braket{\eta_i}{-}}^2>\alpha$, the next step is obtaining the probability that $y=\min\left( \max_{i}\abs{\braket{\eta_i}{+}}^2, \max_{i}\abs{\braket{\eta_i}{-}}^2\right) > \alpha$. In other words, we need the probability that $\max_i \abs{\braket{\eta_i}{+}}^2$ and $\max_i \abs{\braket{\eta_i}{-}}^2$ are simultaneously larger than $\alpha$. We obtain this probability as \begin{equation}\label{eq:DensityDoublet}\begin{split}
P(y > \alpha) &= P\left(\max_i \abs{\braket{\eta_i}{+}}^2>\alpha\right)P\left(\max_i \abs{\braket{\eta_i}{-}}^2>\alpha\right)\\&= \left(  1-\left( I_{\alpha}\left(\frac{1}{2}, \frac{N}{4}-\frac{1}{2} \right)\right)^{N/2} \right)^2.
\end{split}
\end{equation}
The resulting distribution (\ref{eq:DensityDoublet}) suggests that the probability of finding a dominant doublet Hamiltonian $H$ in the centrosymmetric GOE strongly decreases with the system size $N$.\\

\bibliography{Bib_abu_mat2}
\bibliographystyle{apsrev4-1}
\end{document}